\documentclass[fleqn,usenatbib]{mnras}

\usepackage{newtxtext,newtxmath}

\usepackage[T1]{fontenc}

\usepackage{graphicx}	
\usepackage{amsmath}	
\usepackage{tabularx}

\newcommand{\cc}{cm$^{-3}$}
\newcommand{\kmps}{km s$^{-1}$}

\newcommand{\tspec}{T$_{\rm spec}$}
\newcommand{\tew}{T$_{\rm ew}$}
\newcommand{\sigth}{$\sigma_{300}$}

\def \be{\begin{equation}}
\def \ee{\end{equation}}

\title[Hot Gas Emission from Elliptical Galaxies]{X-ray Emission from the Interstellar and Circumgalactic Medium of Elliptical Galaxies based on {\it MACER} simulations}

\author[Vijayan et al.]{
Aditi Vijayan$^{1,2}$,
\thanks{E-mail:aditi.vijayan@anu.edu.au},
Bocheng Zhu$^{1,3}$,
Miao Li$^{4}$
\thanks{E-mail: miaoli@zju.edu.cn},
Feng Yuan$^{1,3}$
\thanks{E-mail: fyuan@shao.ac.cn},
and Luis C. Ho$^{5,6}$
\\
$^{1}$Shanghai Astronomical Observatory, Chinese Academy of Sciences, Shanghai 200030, People’s Republic of China\\
$^{2}$Research School of Astronomy and Astrophysics, Australian National University, Canberra ACT 2601, Australia\\
$^{3}$University of Chinese Academy of Sciences, 19A Yuquan Road, Beijing 100049, People’s Republic of China\\
$^{4}$ Zhejiang University\\ 
$^{5}$ Kavli Institute for Astronomy and Astrophysics, Peking University, Beijing 100871, People's Republic of China\\
$^{6}$ Department of Astronomy, School of Physics, Peking University, Beijing 100871, People's Republic of China\\
}

\date{Accepted XXX. Received YYY; in original form ZZZ}

\pubyear{2015}

\begin{document}
\label{firstpage}
\pagerange{\pageref{firstpage}--\pageref{lastpage}}
\maketitle

\begin{abstract}
Interstellar (ISM) and circumgalactic mediums (CGM) around galaxies are linked to several physical processes that drive galaxy evolution. For example, the X-ray emission from the CGM gas around ellipticals has been linked to the AGN feedback occurring in the host. Upcoming telescopes, such as HUBS with $\sim 2$ eV resolution, can provide us with deep insights about the hot gas properties of such galaxies thus constrain these processes. In this project, we discuss X-ray emission of the ISM and CGM of elliptical galaxies simulated using {\it MACER} code. We generate X-ray emission data from the {\it MACER} simulations with various feedback models and produce mock observations for an instrument with high spectral resolution, which is a  necessary step of selecting sources for the future observations with planned mission such as HUBS. More importantly, we establish connections between the physics of AGN and stellar feedback  with the emission spectra from the ISM and CGM  to investigate the possibility of using observations to constrain feedback models. We fit the X-ray spectra from these simulations with standard fitting procedures and compare the retrieved physical properties with their counterparts from the simulations to understand whether the future high-resolution observations can reliably reveal the properties of the gas in the galaxies.
\end{abstract}




\section{Introduction}

Feedback processes occurring in a galaxy as a result of the activity of the central Active Galactic Nucleus (AGN) are known to affect the evolution of the host galaxy. These processes are driven by the interaction  of the radiation, wind and jet outputs from the AGN with the gas in the galaxy. Energetically, a large fraction ($\sim 10\%$) of the mass accreted onto the supermassive black hole (SMBH) is converted into energy for powering the AGN. Because this energy is significantly larger than the binding energy of gas in the galaxy, an AGN strongly impacts the gas distribution host galaxy \citep{Fabian2012}.

Several observations indicate the effect an AGN has on its host. For relatively larger objects, such as clusters and groups, scalings between the black hole mass (M$_{\bullet}$) and the halo temperatures are explained only by invoking AGN feedback as a source of additional heating to the gas \citep{Bogdan+18, Gaspari+19, Lakhchaura+19}. In early type galaxies (ETGs), which are the focus of this paper, observations have established a steep and tight correlation between the total halo X-ray luminosity (L$_{\rm X}$) and X-ray temperature, T$_{\rm X}$,  which is hard to be explained using the self similarity argument alone \citep{Boroson+11, Kim&Fabbiano13, Kim&Fabbiano15, Babyk+18}. Further, it has been established that hot atmospheres of ETGs cannot be created through the action of stellar feedback alone \citep{Goulding+16, Werner+19}. These clues suggest that AGN feedback plays significant role in moulding the properties of hot atmospheres around ETGs.

In numerical simulations, the AGN feedback is implemented via two modes. The nomenclature for these modes is diverse in the literature. Throughout this work, we refer to them as ``cold'' and ``hot'' modes, following \cite{Yuan+18}. Cold mode, also referred to as the ``quasar mode'' or ``radiative mode'', operates in AGNs which are powered by cold accretion flow when the accretion rate is above $\sim 2\%$ Eddington rate ($\dot{M}_{\rm Edd}\equiv 10L_{\rm Edd}/c^2$). The main outputs in this mode are radiation and wind \citep{Morganti2017,Yuan+18}.  

The hot mode, also the kinetic or jet or radio or mechanical mode, operates in AGNs which are powered by hot accretion flow when the mass accretion rate is below $\sim 2\% \dot{M}_{\rm Edd}$. The outputs of the AGN in this mode are radiation, wind, and jet \citep{Yuan+18}. It has been suggested that wind launched in this mode may prevent the gaseous atmosphere from cooling and forming stars, thereby maintaining the quiescent state of the galaxy \citep[also Zhu et al 2022 (in prep) for details]{Yao+21}, or even inducing quenching of the galaxy \citep{2017MNRAS.465.3291W}. Observationally, this mode has been identified in clusters from the presence of a cavity in the X-ray emission which is filled with radio emission \citep[see for a review]{Werner+19}. 

Unfortunately, works focussed on simulating AGN feedback simulation are quite diverse in the sense that different works adopt different models and not all the above-mentioned AGN physics has been properly incorporated \citep{Ostriker2017}. \cite{Yoon+19} shows that physical properties of the system such as the kinetic power output of the AGN, which determines the mass of gas pushed out from the galaxy, depend on the feedback model adopted. 
As such the feedback model adopted determines the temporal and spatial distribution of gas density and temperature \citep{Yuan+18}, which in turn characterise the observational properties of the galaxy. Some simulations have successfully reproduce key observations such as the presence of buoyant cavities \citep{Gaspari+12, Gaspari+14}, M$_{\bullet}$-$\sigma$ relation (\cite{Sijacki+07, Booth&Schaye11, Teyssier+11, Choi+15}, etc). In this project, we tackle one aspect where tension exists between simulations and observations, i.e., diffuse X-ray emission of ETGs.


Diffuse soft X-ray emission ($\lesssim 2$ keV) originates from hot gas within and around a galaxy. Such a diffuse emission has been observed and studied around star-forming galaxies \citep{Yamasaki+09, Anderson+11, Dai+12, Anderson+16, Bogdan+13, Bogdan+13B, Bogdan+17, Lopez+20} as well as more massive galaxies \citep{Anderson+15, Kim&Fabbiano15}. The emission is usually characterised by the total X-ray luminosity  (L$_{\rm X}$) and the temperature of the emitting gas, T$_{\rm X}$, estimated by fitting the X-ray spectrum by emission models. Simulations of early-type galaxies (ETGs) \citep{Gaspari+12, Gaspari+14} report a lower than expected L$_{\rm X}$ and a break in the L$_{\rm X}$-T$_{\rm X}$ relation neither of which are corroborated by observations \citep{Babyk+18}. \cite{Choi+15} analysed separately the effects of the two feedback mechanisms and found that their quasar (thermal) mode of feedback overestimates L$_{\rm X}$ by nearly two orders of magnitude. Cosmological simulations such as Illustris-TNG \citep{Truong+20} and EAGLE \citep{Schaye+15} also report brighter than observed X-ray emission as a result of their respective AGN feedback recipes \citep{Davies+19}.

Apart from global X-ray properties such as L$_{\rm X}$, the X-ray emission spectrum carries a wealth of information about the temperature and the chemical abundance of the hot emitting gas. Usually, spectral fitting models, based on simplifying assumptions about the underlying emitting plasma, are used to derive the temperature of the hot gas from the spectrum (\tspec). Studies of simulated emission from clusters have found that the spectra-derived temperature T$_{\rm spec}$ may differ from emission-weighted temperature (\tew), leading to underestimation in the latter of up to $\sim 20\%$ \citep{Mazzotta+04, Rasia+05, Vikhlinin06}. Spectrum-derived temperature estimates of diffuse gas around galaxies are also dependent on whether the fitting models use a single or multi-temperature components \citep{Li&Wang13, Wu+20}. 

The current observations of diffuse X-ray emission around ETGs are limited in scope because their poor spectral resolution and sensitivity, and the small field of view of the present telescopes are not attuned to study low density gas ($\sim 10^{-2}$ \cc at $\sim 10$ kpc) which fills the region around ETGs. In this context, our focus in this work is the upcoming X-ray mission Hot Universe Baryon Survey (HUBS) which is specifically designed to study the hot, diffuse gas around galaxies, groups and cluster \citep{Cui+20}. With respect to AGN feedback occurring in ETGs, this presents an opportunity to study the observational effects of the hot and cold feedback mechanisms on the circumgalactic medium (CGM) around such galaxies. The high spectra resolution of HUBS ($\sim 2$ eV, compared to $\sim 130$ eV of \textit{Chandra}) will allow for accurate temperature measurements through fitting of emission lines \citep{Bohringer&Werner10}. Further, a comparison with observations can indicate how well various AGN feedback recipes, implemented in different simulations, emulate the physical processes occurring in nature and whether some of the discrepancies with respect to observation can be removed. 

In this project, we aim to understand the X-ray properties of the ISM within and the CGM around isolated ellipitcal galaxies. Specifically, we analyse the relationship between an estimate of X-ray temperature, obtained from the emission spectrum, and the physical temperature of the gas in the galaxy. We analyse results from high-resolution simulations presented in \cite{Yuan+18}, which follow the evolution of an isolated elliptical galaxy (i.e., not considering galaxy merger and cosmological inflow to the galaxy) based on the {\it MACER} code. The two main features of the model are that: 1) the black hole accretion rate is precisely determined because the inner boundary of the simulation and is typically ten times smaller than the Bondi radius of the accretion flow and, 2) the state-of-the-art AGN physics is incorporated into the code, including the wind and radiation as a function of mass accretion rate. We briefly introduce the key components of the model in Section \ref{sec:models}. 
The main objective of this paper are- (i) to understand how the density and temperature distributions of gas changes with the nature of feedback implemented in MACER. (Section \ref{sec:density_temperature_distribution}); (ii) to understand whether the differences in the feedback lead to differences in the emission spectra from the ISM and the CGM. (Section \ref{sec:extract_pure_spectra}); (iii) to study the differences between the theoretical (\tew) and observational properties (\tspec) of the galaxies. (Section \ref{sec:extract_pure_spectra}, Figures \ref{fig:tspec_tew_0-10kpc} and \ref{fig:tspec_tew_10-300kpc}); (iv) to understand what causes the discrepancy between these properties. (Section \ref{sec:compare_tspec_tew}); and (v) to find out whether we can use spectral differences and radially averaged emission to discriminate between feedback recipes. (Section \ref{sec:Mock_Image}).


\section{Models}\label{sec:models}

In this section we give a brief summary of the {\it MACER} simulations presented in \citet{Yuan+18}. Readers are also referred to \citet{Yuan+20} for an overview of the {\it MACER} code. {\it MACER} are a set of 2D, axi-symmetric simulations which focus on the evolution of a single elliptical galaxy, with the inner and outer boundaries being $\sim 2~{\rm pc}$ and $500~ {\rm kpc}$, respectively. A resolution as high as $\sim 0.3~{\rm pc}$ is achieved at the inner boundary. The Bondi radius of the accretion flow is typically $\sim 15~ {\rm pc}$ \citep{Yao+21}, which is several times larger than the inner boundary of the simulation, thus the outer boundary of the accretion flow is well resolved. Once the accretion rate at the inner boundary of our simulation domain is calculated, we can safely combine the accretion physics as subgrid model and precisely calculate the mass accretion rate of the AGN at the black hole horizon. This is crucial to determine the strength of the AGN feedback. 

According to the value of the mass accretion rate, the accretion is divided into ``cold'' and ``hot'' modes, bounded by $\sim 2\% L_{\rm Edd}$ \citep{Yuan2014}. Radiation and wind are present in both modes while jet is perhaps  present only in the hot mode\footnote{Maybe jets are also present in some cases when the accretion is in the cold mode, hinted by the existence of radio-loud quasars. This is still an unsolved problem. Jet has not been included in the hot mode in \citet{Yuan+18} and it is being added into the code in Guo et al. (2022, in preparation).}. The properties of the wind in the cold mode, including the velocity and mass flux as a function of AGN luminosity, are taken from observations \citep{2015MNRAS.451.4169G}. Wind in the hot mode has been intensively studied in recent years by magnetohydrodynamical numerical simulations \citep{Yuan2012,Narayan2012,Yuan2015,Yang2021}. In the observational side, we are accumulating more and more observational evidences for wind from hot accretion flows, including the supermassive black hole in our Galactic center \citep{Wang2013,Ma2019}, low-luminosity AGNs \citep{Cheung2016,2019ApJ...871..257P,Shi2021,2022ApJ...926..209S}. Especially, \citet{Shi2021,2022ApJ...926..209S} have found the most direct evidences for winds in two prototype low-luminosity AGNs by detecting the blue-shifted emission lines. But still, we have not obtained good constrain to the hot wind properties, therefore we adopt the properties of wind from GRMHD simulations of \citet{Yuan2015}. 

Wind and radiation are injected at the inner boundary of the simulation domain as the boundary conditions. Their energy and momentum interaction with the gas in the host galaxy is calculated self-consistently in the code. In addition to AGN feedback, the code also includes star formation and stellar feedback. We follow the evolution of the galaxy for $\sim 2.5$ Gyr up to the present redshift. 

We consider four different models in the present work, as listed in Table \ref{tab:model}, with different AGN feedback models and galaxy properties. The first ``Fiducial'' model is  directly taken from \citet{Yuan+18}, in which the AGN feedback physics as mentioned above has been properly incorporated. Specifically, both ``hot'' and ``cold'' modes have been taken into account according to their accretion rates. In the ``Only Cold'' (``Only Hot'') model, irrespective of the accretion rate, we always adopt the physics of the cold (hot) mode for the descriptions of wind and radiation.  In the ``$\sigma_{300}$'' model, the AGN physics is identical to the Fiducial model, but we change the the value of the velocity dispersion of the galaxy from $200~ {\rm km~s^{-1}}$ to $300~ {\rm km~s^{-1}}$. Using these four models, we mimic the effects of different AGN feedback physics and galaxy size. One main caveat of these models is that we do not consider the effects of cosmological inflow. It is expected that it will affect the physical properties of the CGM of the galaxy. We will consider this effect in the future work.

\begin{table*}
\caption{Model description}
\begin{center}
\begin{tabular}{|c|c|}
\hline
     Name & Features  \\ 
      \hline
      Fiducial & Identical to the one described in \cite{Yuan+18}.\\
      \hline
      OnlyCold & Adopting the cold-mode AGN physics no matter what the value of the accretion rate.\\
      \hline
      OnlyHot & Adopting the hot-mode AGN physics no matter what the value of the accretion rate.\\
      \hline
      $\sigma_{300}$ & The galaxy velocity dispersion is set to $300$ \kmps as compared to $200$ \kmps in the Fiducial run.\\
      \hline
\end{tabular}
\end{center}\label{tab:model}
\end{table*}

\section{Results}\label{sec:results}

\begin{figure}
	\includegraphics[width=\columnwidth]{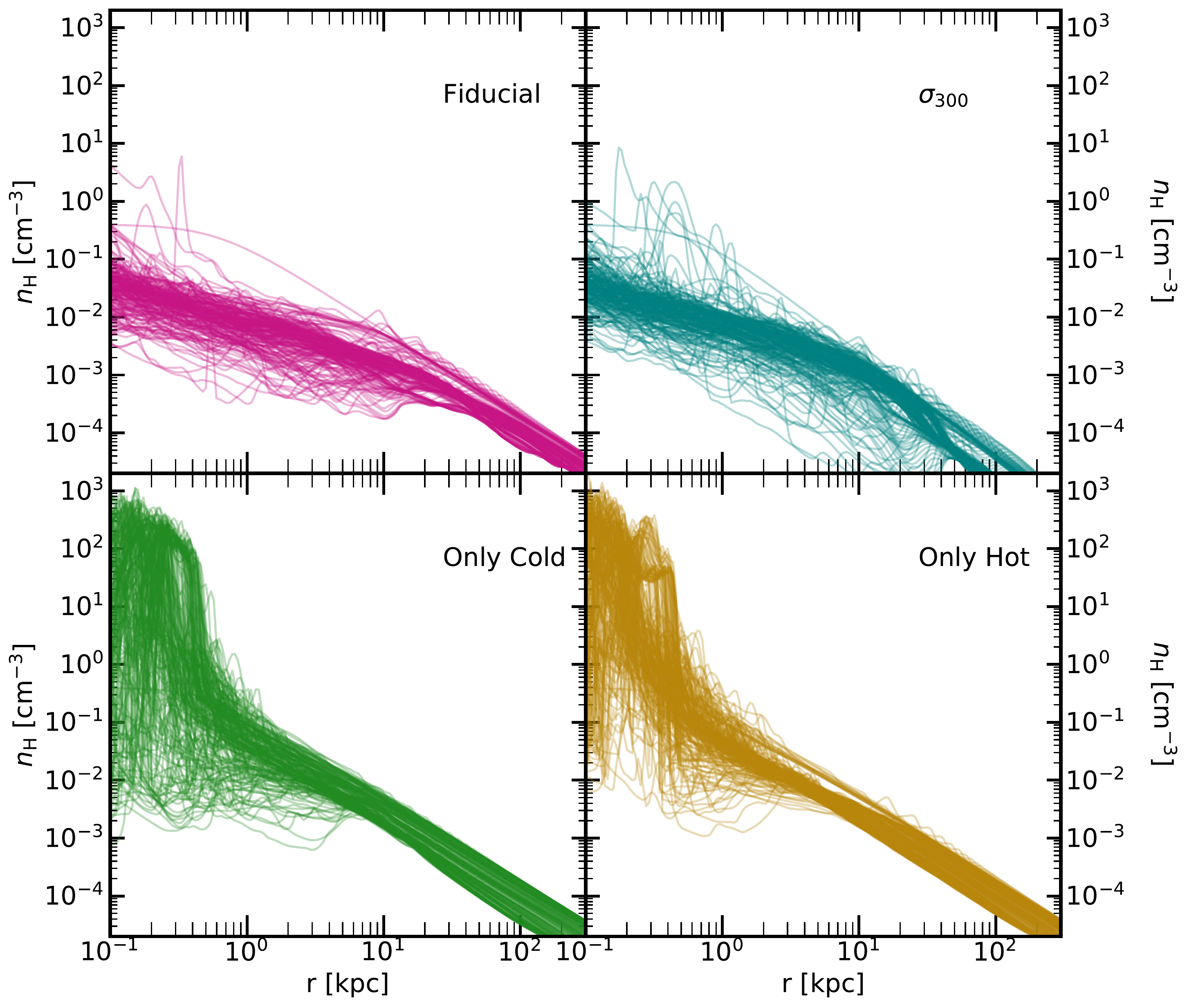}
    \caption{The radial variation of number density ($n_{\rm H}$) for the four models. Each curve represents a different time in the simulation.}
    \label{fig:nH_vs_rad}
\end{figure}

\begin{figure}
	\includegraphics[width=\columnwidth]{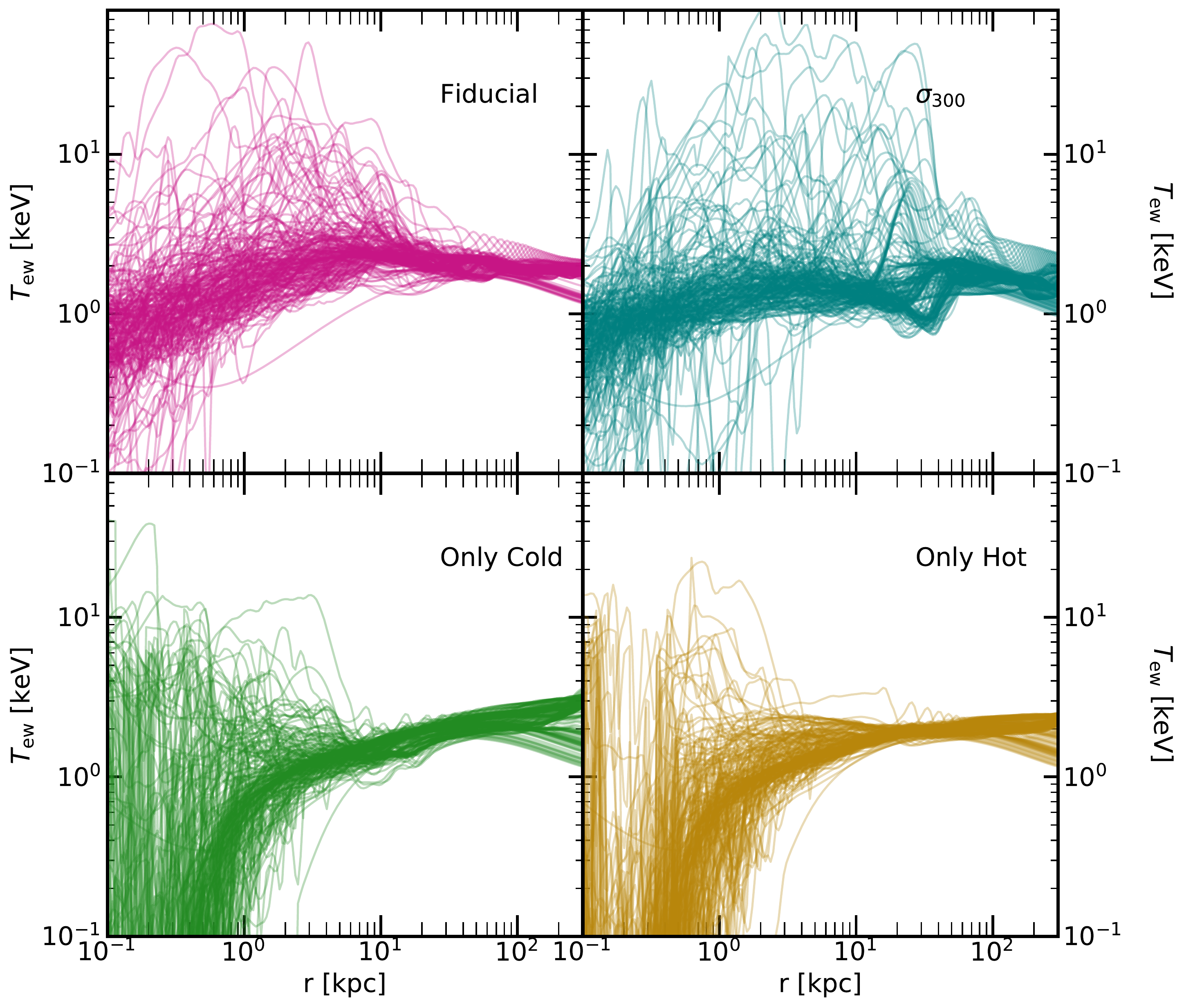}
    \caption{Identical to Figure \ref{fig:nH_vs_rad}, but instead of number density, we show the emission-weighted temperature, defined by Equation \ref{eqn:tew}.}
    \label{fig:tew_vs_rad}
\end{figure}

\begin{figure}
	\includegraphics[width=\columnwidth]{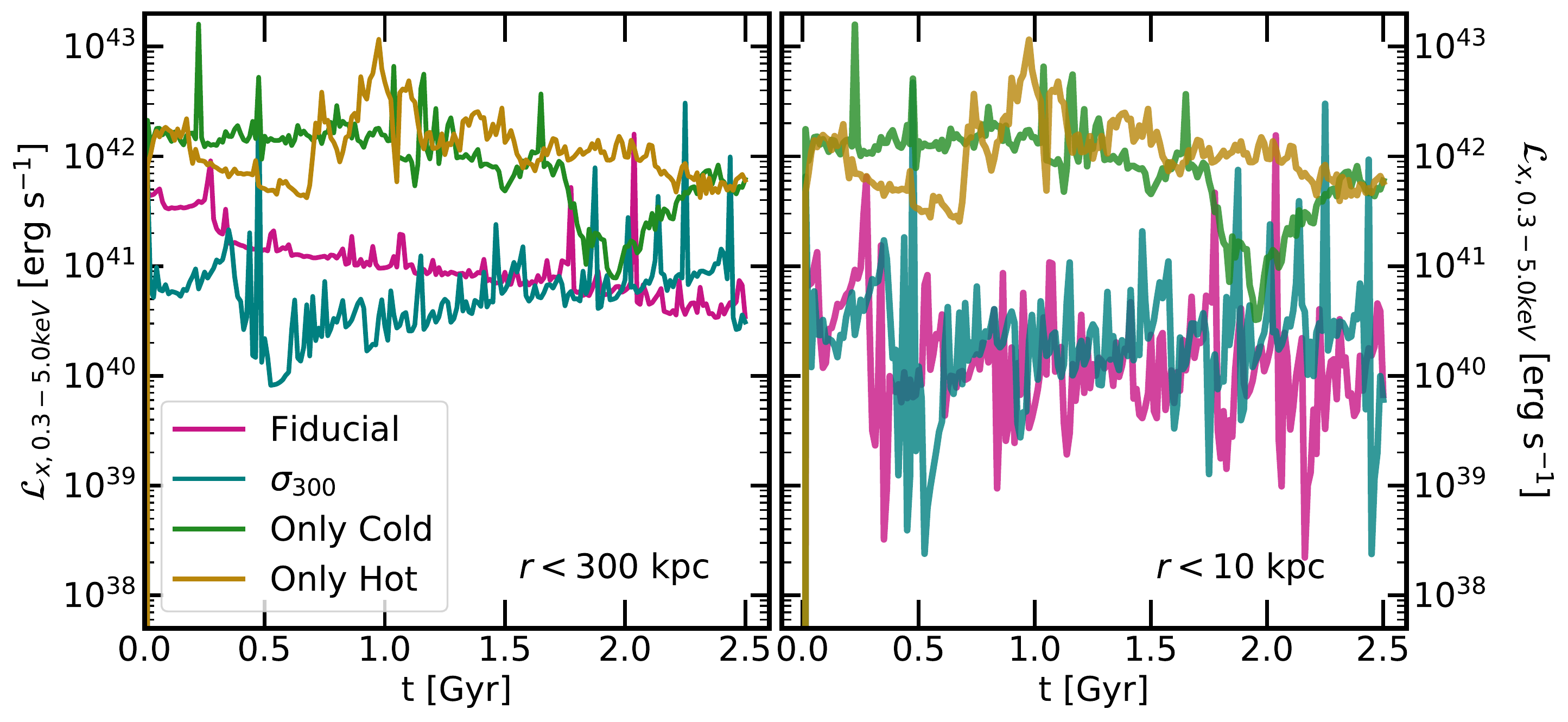}
    \caption{Temporal variation of the total soft ($0.3-5.0$ keV) X-ray luminosity for the four runs, calculated for $r<300$ and $r<10$ kpc. }
    \label{fig:lx_vs_t}
\end{figure}

\subsection{Density and Temperature Distribution}\label{sec:density_temperature_distribution}

We first discuss the temperature and density distributions of gas in different models. Figures \ref{fig:nH_vs_rad} and \ref{fig:tew_vs_rad} show the radial distribution of density and the emission-weighted temperature in the four models. Each curve in every panel represents one time in the simulation. 

As expected, the density profiles are declining outward radially. Upto $r>200$ kpc, the profiles are similar to those following the hydrostatic equilibrium. The profiles for ``Only Hot'' and ``Only Cold'' models are flat within $1$ kpc and drop by five orders of magnitude beyond this radius. The ``Fiducial'' and ``\sigth\ '' have monotonically declining profiles up to $r\gtrsim 200$ kpc. We also note that the density profiles in the inner region ($r<10$ kpc) of the ``Only Hot'' and ``Only Cold'' modes vary significantly in time, up to nearly four orders of magnitude. In the outer regions, $r>10$ kpc, this feature is reversed, that is, in the ``Only Hot'' and ``Only Cold'' models there is hardly any variation in density profiles, while for the "Fiducial'' and ``\sigth'', there is a greater variation.

We define the emission-weighted temperature as 
\be\label{eqn:tew}
T_{\rm ew} = \frac{\int n^2 \Lambda(T) T dV}{\int n^2 \Lambda(T)  dV}\,,
\ee
where $\Lambda(T)$ is the X-ray emissivity between $0.3-5.0$ keV\footnote{This energy range is for estimating the emission-weighted temperature for \textit{Chandra}-like low spectral resolution instrument.}, $T$ is the gas temperature and $n$ is the gas number density. The radial profiles for four different models are shown in Figure \ref{fig:tew_vs_rad}. As for Figure \ref{fig:nH_vs_rad}, each curve represents a different time step in the evolution. Unlike the density profiles, the temperature profiles have a vary across the radial extent, from $0.1-3$ keV. 

The general trends of density and temperature of the Fiducial and $\sigma_{300}$ models shown in the above two figures are not difficult to understand as they are similar to profile representing hydrostatic equilibrium. However, for ``Only Cold'' and ``Only Hot'' models, we can see that the density (temperature) rapidly increases (decreases) inward within $\sim 1 {\rm kpc}$. As shown in \citet{Yuan+18}, the main energy input from the AGN is by wind. The significantly smaller temperature (and thus larger density) in ``Only Cold'' and ``Only Hot''  models compared to the Fiducial model is because the energy input from wind is much weaker in the former two cases. For the ``Only Cold'' model, compared to the Fiducial model, the wind power remains same when the accretion is in the cold mode. But when the accretion rate is low and the accretion is in the hot mode, the wind described by the ``cold-mode physics'' will be weaker than that described by the ``hot-mode physics''. Similarly, for the ``Only hot'' model, the wind power remains same when the accretion rate is low. But when the accretion rate is high, the wind described by the ``hot-mode physics'' will be weaker than that described by the ``cold-mode physics''. 

The density and temperature distribution of the gas determines the X-ray luminosity of the system. We show the soft X-ray luminosity integrated from the whole galaxy plotted against the simulation time in Figure \ref{fig:lx_vs_t}. To obtain the luminosity from the simulation data, we use APEC emissivity tables between $0.3$ and $5.0$ keV. We show the luminosity from the entire simulation domain (left panel) as well as that from the ISM, $r<10$ kpc. For all the models, the stochasticity in the density and temperature profiles is translated into the temporal variations in the luminosity.

Given the high stochasticity of the density and temperature profiles of the ``Only Hot'' and ``Only Cold'' models in the ISM, there is significant variation in the total luminosity from this region. Further, these models have higher luminosity because on an average they have higher density values, especially in the ISM of the galaxy.

\subsection{Extracting Pure Emission Spectra}\label{sec:extract_pure_spectra}

\begin{figure}
	\includegraphics[width=\columnwidth]{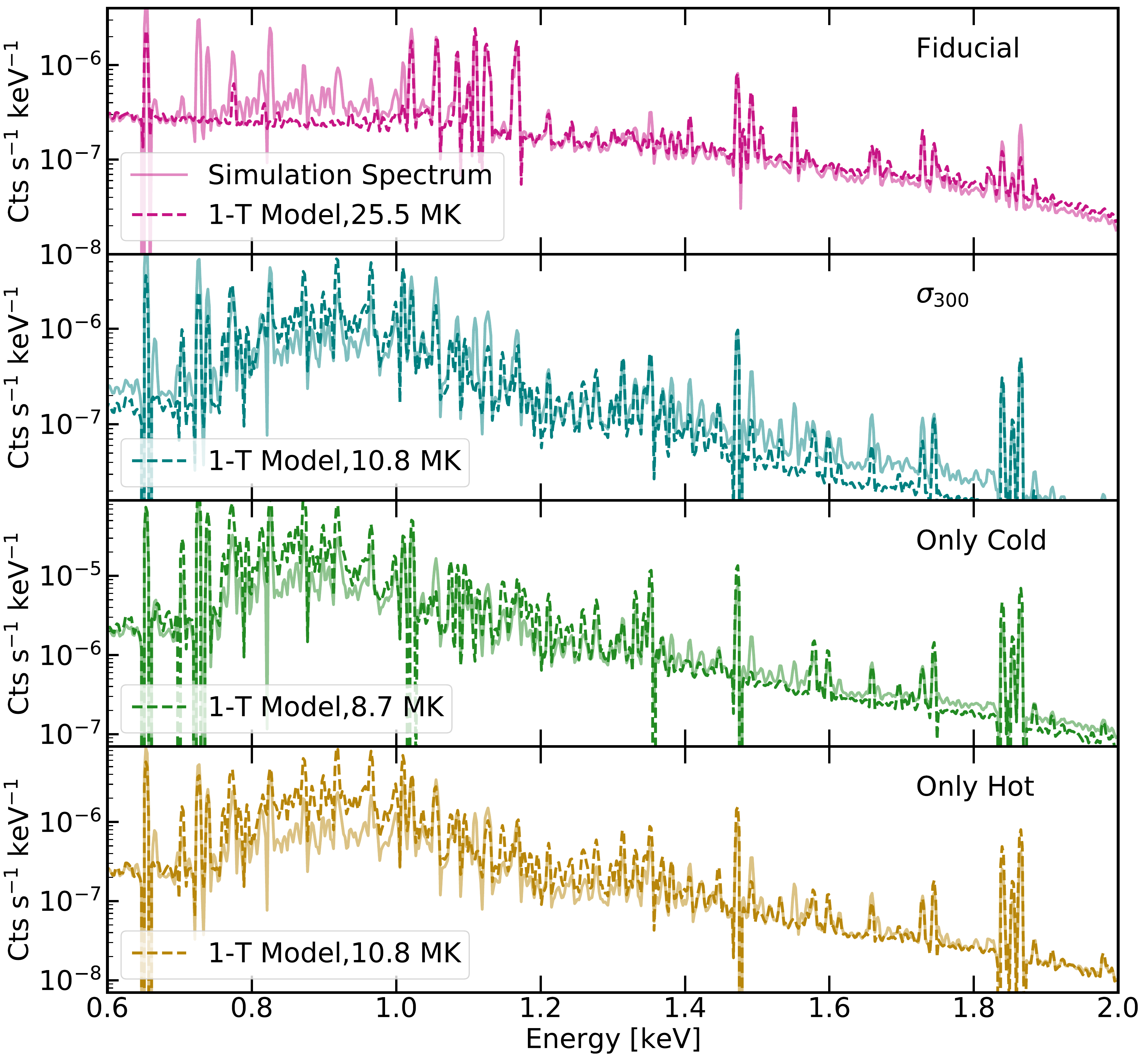}
    \caption{The simulation spectra after $1.25$ Gyr of evolution, which roughly marks the midway of the simulation for the different AGN models, and their comparison with the spectra produced by the 1-T model. The solid lines represent the simulation spectra, generated using pyAtomDB, as described in the text. The best-fit spectra are shown by the dashed curve. The spectra are generated using the gas between $0.1-10.0$ kpc, that is the ISM of the galaxy. The number indicates the fitting temperature, T$_{\rm spec}$.}
    \label{fig:sim_spec}
\end{figure}

We use pyatomDB \citep{Foster+20} to generate a spectral emission from a parcel of gas over a wide range of temperature, density and metallicity. pyatomDB \citep{Heuer+21} is a database for modelling the X-ray emission from collisionally-ionised plasma. The users have the option to convolve the spectra generated with an instrument response to produce a realistic spectrum. In this Section, we discuss the spectrum derived temperature, \tspec, low- and high-spectral resolution instruments.

For our analysis, we separate the simulation domain into ISM ($r<10$ kpc) and CGM ($10<r<300$ kpc) regions. From Figure \ref{fig:tew_vs_rad}, we note that for all the models, the temperature varies dramatically within the ISM, while it is nearly flat (for all expect \sigth\ model) in the CGM. Given these profiles, we expect the spectra to be similar in the CGM region of the galaxy since the X-ray emission spectrum depend on the plasma temperature. We use the density, temperature and metallicity information from each cell in the ISM region ($r<10$ kpc) of the simulation domain as an input for pyatomDB and generate a spectrum for every cell. For low resolution \tspec, we set the instrument response in the pyatomDB session corresponding to the instrument AISCC (on the \textit{Chandra} telescope having a resolution of $\sim 130$ eV) and for the high resolution \tspec, we use the response for HUBS telescope\footnote{HUBS is an upcoming X-ray telescope designed specifically to observe hot gas around galaxies, having a $1^{\circ} \times 1^{\circ}$ FoV and a spectral resolution of $2$ eV \citep{Cui+20}. The energy range of HUBS is $0.5-2.0$ keV, therefore, we show the spectra in this energy range.}. We then add the spectrum emitted by each cell and sum it up to produce a single spectrum for the entire simulation domain.  Figure \ref{fig:sim_spec} shows the high-resolution spectra for the four models, after $1.25$ Gyr of evolution, in dashed lines. As expected, the emission decreases with increasing energy.

We extract the temperature of the X-ray emitting gas from the spectra, which is an observable quantity obtained by fitting models to the observed spectrum. We use the recipe followed by observers to extract a spectral temperature, T$_{\rm spec}$. We fit the spectra using a 1-T fitting model similar to the fitting methods described in \citep{Truong+20}, but with some minor modifications described below. The 1-T model assumes that the observed spectrum is the result of emission from gas at a single temperature. We first generate a set of ``ideal'' 1-T spectra for temperatures in the range of $10^4-10^9$ K, corresponding to the temperature range of the X-ray emitting gas in the simulated galaxy. For each of these temperatures, we vary the abundance parameter in the pyAtomDB session between $10^{-3}-1.5$ Z$_{\odot}$.
For every timestep, we compare the simulation spectrum over the entire set of the ``ideal'' 1-T spectrum generated using pyAtomDB.
To estimate the best-fit temperature, we minimized a statistical quantity, called the Wasserstein distance \citep{Rubner+98}, between the simulation spectrum and the ideal spectrum generated at that temperature. The temperature corresponding to the minimum distance is taken to be the best-fit temperature, which we denote as T$_{\rm spec}$. We also note here that the spectra- both ideal and simulation- do not include foreground and background contributions.

In Figure \ref{fig:sim_spec}, we show the simulation spectrum from the ISM as well as the corresponding spectrum from the best-fit single temperature model after $\sim 1.25$ Gyr of evolution. The solid curves in each panel represent the spectrum from the simulations, while the dashed curves are the best-fit spectra, obtained using the procedure outlined above, with the fitting temperature in the bottom left corner. The spectra obtained from 1-T model provide a reasonable fit for the simulation data, even though it is quite simplistic. However, it is not a perfect fit as it misses out, for e.g., low-temperature emission ($<1.0$ keV) in Fiducial model and relatively high energy emission ($>1.6$ keV) in the \sigth\ model.
We note that the spectra corresponding to \sigth, ``Only Cold'' and ``Only Hot'' models show similar hump-like feature around $0.8-1.0$ keV, while the Fiducial run does not. Note that the fitting temperatures are different for all four models.

\subsection{Comparing Spectral Temperature with Emission-Weighted Temperature}\label{sec:compare_tspec_tew}

\begin{figure}
	\includegraphics[width=\columnwidth]{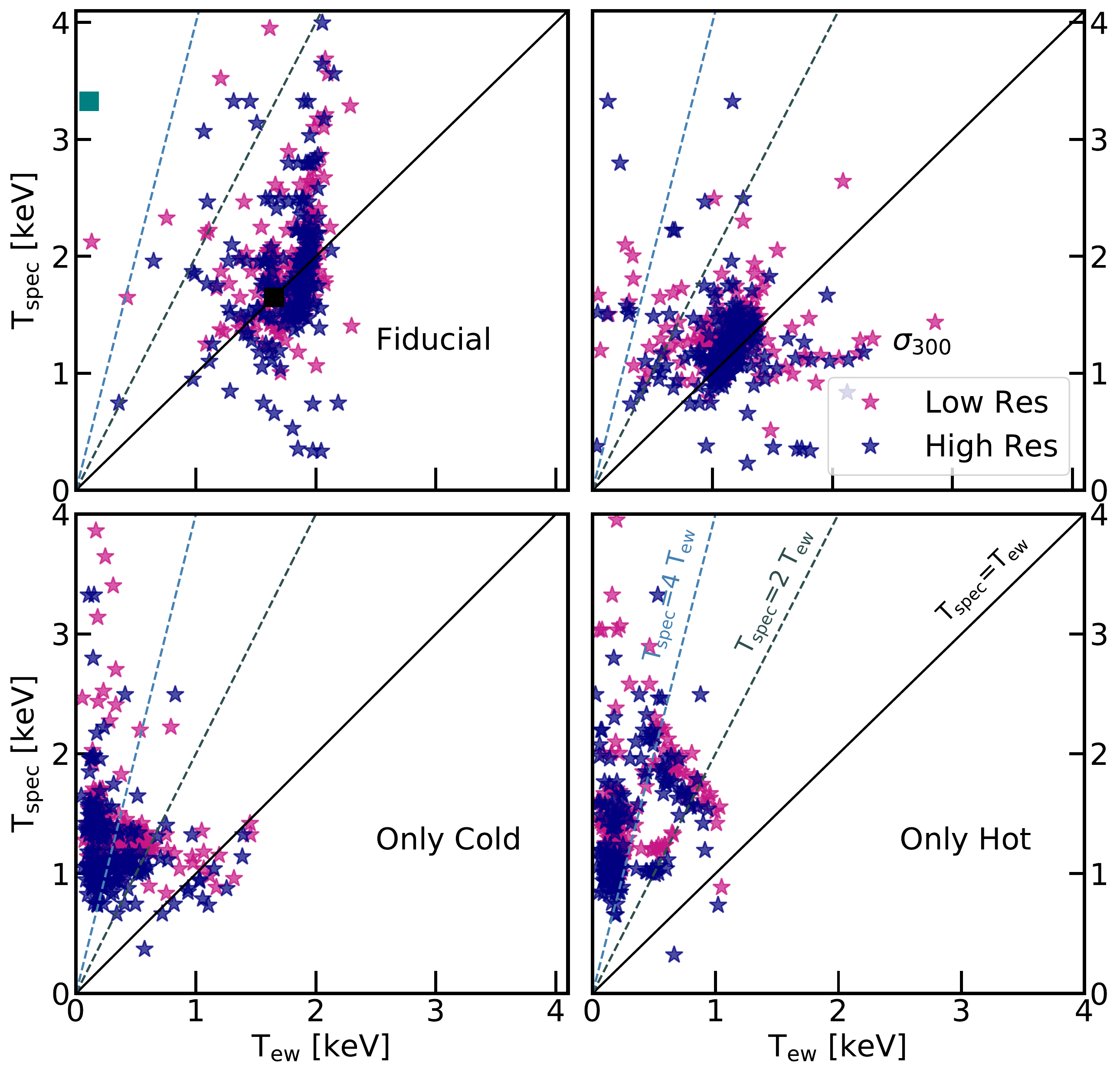}
    \caption{The distribution of T$_{\rm spec}$ with the emission-weighted temperature, T$_{\rm ew}$, indicating the discrepancy between the two quantities for various models. Each star corresponds to a different time step in the simulation. We show \tspec\ generated for low ($\sim 130$ eV) and high ($\sim 2$ eV) spectral resolution instruments. For reference, we show curve corresponding to 
    \tspec$=$\tew (black, solid), $=2$\tspec (grey, dashed) and $=4$\tew (blue, dashed). We have used the region between $0.1-10$ kpc for estimating the two temperatures. The black and teal points indicate time steps for which \tspec$>$ and \tspec$\sim$\tew, respectively. We show the luminosity-weighted temperature-density histogram for these time steps in Figure \ref{fig:histo_lx}. Though there are time steps in all the models, for which \tspec\ accurately predicts \tew, it is not always the case. }
    \label{fig:tspec_tew_0-10kpc}
\end{figure}

\begin{figure}
	\includegraphics[width=\columnwidth]{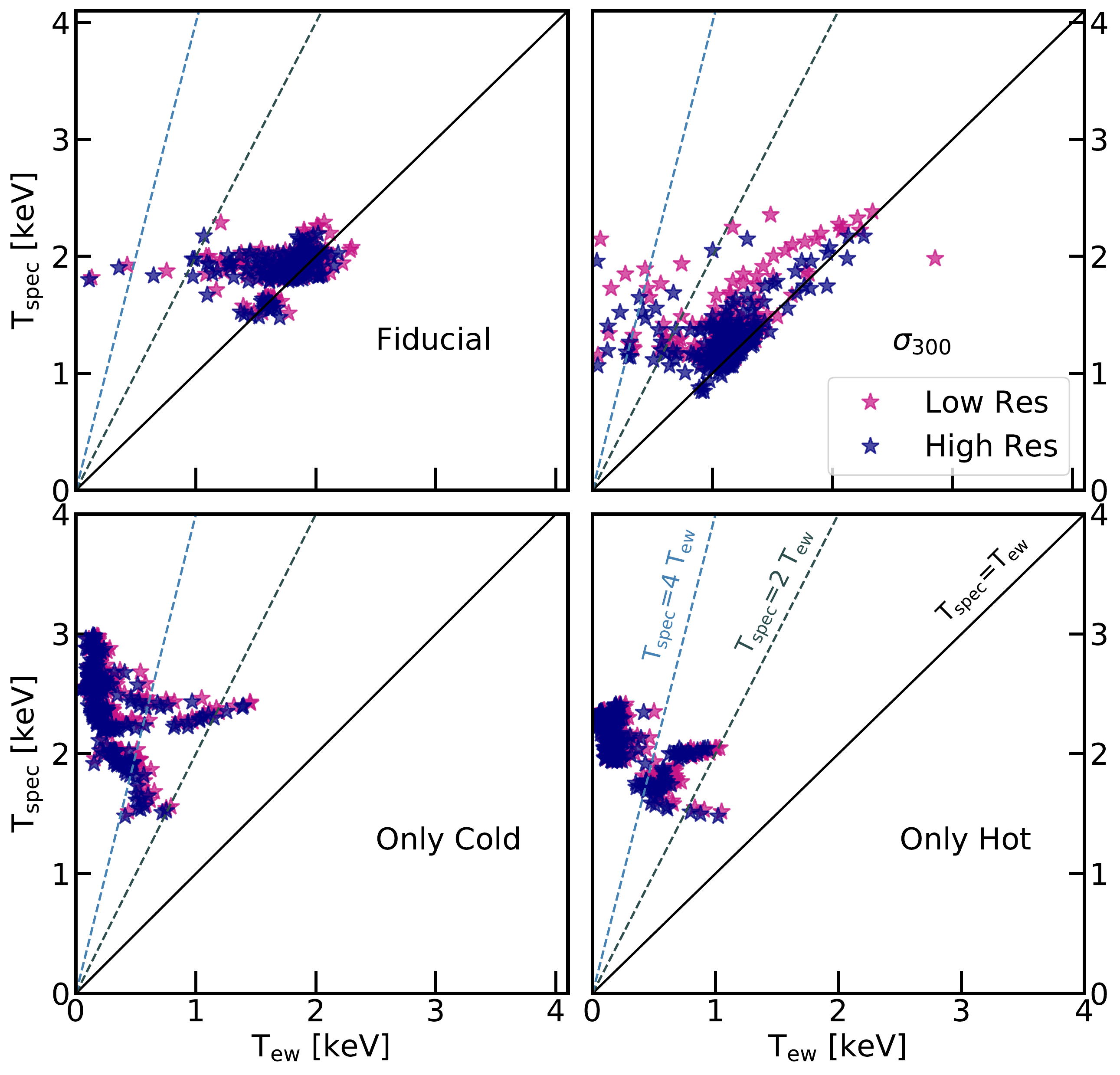}
    \caption{Identical to Figure \ref{fig:tspec_tew_0-10kpc}, but for the region between $10-300$ kpc.}
    \label{fig:tspec_tew_10-300kpc}
\end{figure}

\begin{figure}
	\includegraphics[width=\columnwidth]{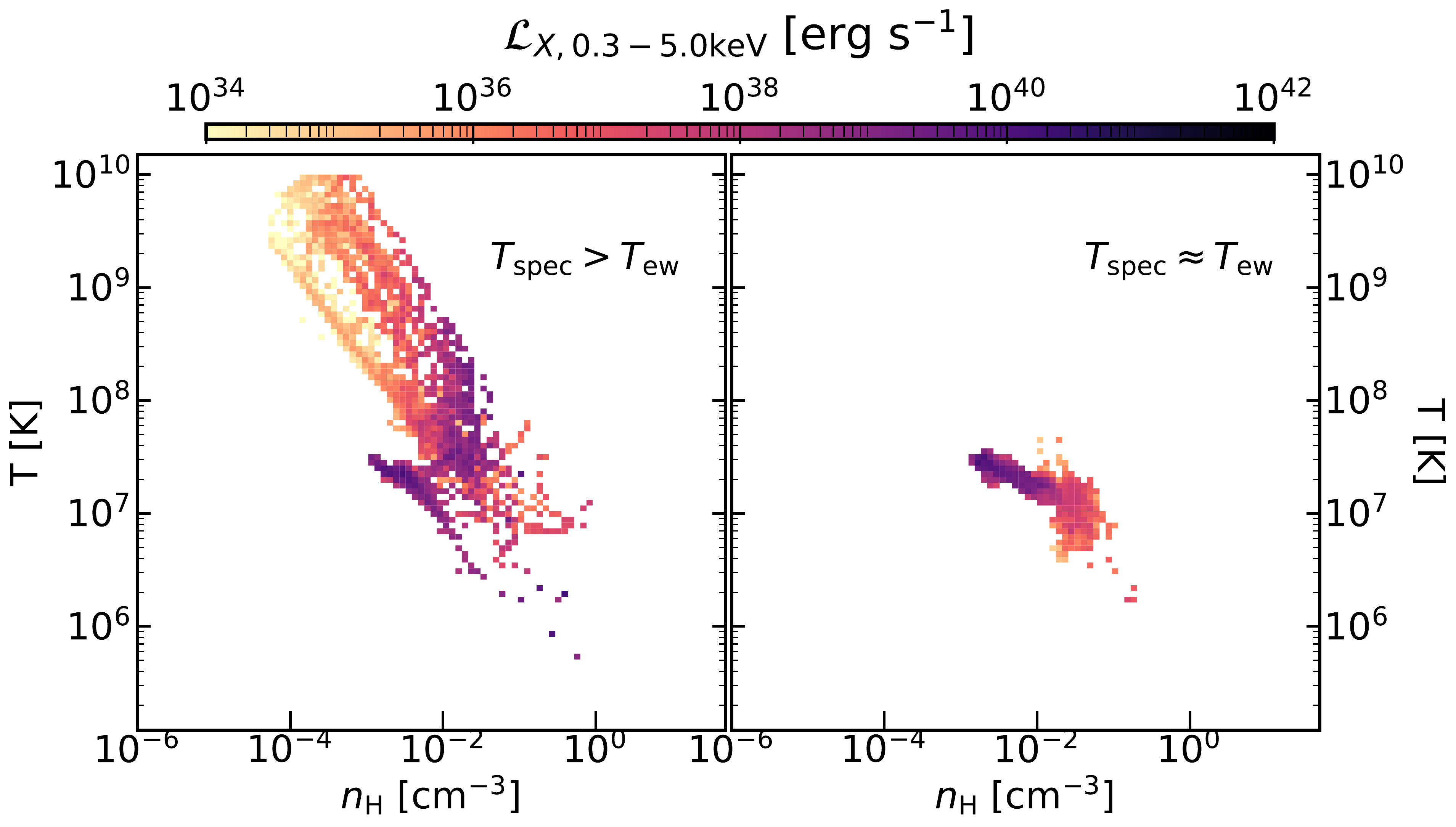}
    \caption{The temperature-density distribution of luminosity for two time steps of the Fiducial run. The left and right panel correspond to the teal and black squares in Figure \ref{fig:tspec_tew_0-10kpc}, respectively.}
    \label{fig:histo_lx}
\end{figure}

The physical interpretation of T$_{\rm spec}$ is that it represents the emission-weighted temperature of the gas. To assess how accurate this interpretation is, we compare the T$_{\rm spec}$ with the emission-weighted temperature, T$_{\rm ew}$, calculated using Equation \ref{eqn:tew}. From the simulation data, we know the emission-weighted temperature exactly. For each time step, we now have a pair of temperatures- T$_{\rm ew}$ and T$_{\rm spec}$. Though both temperatures are obtained from the simulation data, T$_{\rm spec}$ is the temperature that derived from spectra, while T$_{\rm ew}$ is the \textit{actual} temperature of the gas. Such a comparison, between these two different temperatures, is not possible with observations.

Figures \ref{fig:tspec_tew_0-10kpc} and \ref{fig:tspec_tew_10-300kpc} show scatter plots between T$_{\rm ew}$ and T$_{\rm spec}$ for the ISM and the CGM ($r>10$ kpc), respectively. The spectrum from the ISM (CGM) is obtained by including only those cells that lie within (without) $10$ kpc. \footnote{Actual spectra from the ISM region will possess contribution from the intervening gas between the observer and the CGM. However, from the radial surface brightness plot (Figure \ref{fig:rad_sb}) we estimate that this contribution will be negligible and therefore, we ignore it for our analysis.}  We obtain \tspec\ from the spectra for high-resolution (magenta points) and low-resolution (navy points) instruments, for several time steps. If the two temperatures are identical, they should fall on the solid black line which represents the equality T$_{\rm spec}$=T$_{\rm ew}$. In the figures, we also show lines corresponding to $2$ T$_{\rm ew}$ and $4$ T$_{\rm ew}$. While there exist some time steps when the predicted T$_{\rm spec}$ is equal or close to the physical T$_{\rm ew}$, the equality does not hold true for most time steps during the simulation. Overall, we note that for both inner- and outer-CGM, the predicted \tspec\ overestimates \tew\ by a factor of $\gtrsim 4$.

There are considerable differences in the \tspec-\tew\ distribution between the ISM and CGM. For the CGM, the predicted \tspec\ by the lower resolution instrument are close to their higher resolution counterpart, as the navy and magenta points are overlapping for several time steps, while for ISM the two predict different \tspec. 

To understand the source of the discrepancy between \tspec\ and \tew, we show the luminosity-weighted temperature-density histogram of two time steps in Figure \ref{fig:histo_lx}. We have chosen these particular time steps from Figure \ref{fig:tspec_tew_0-10kpc} where these are shown in black and teal squares in the panel for the Fiducial run. The teal square corresponds to a time step for which \tspec$>$\tew\ at $0.275$ Gyr and for the black square at $1.64$ Gyr the two temperatures are nearly identical. For Figure \ref{fig:histo_lx}, we distribute the gas in $0.1-10$ kpc in several temperature and density bins and weighted using the soft X-ray luminosity. The gas distribution corresponding to the time step for which T$_{\rm spec}>$ T$_{\rm ew}$ (teal point in Figure \ref{fig:tspec_tew_0-10kpc}) has a wide temperature range ($10^3 \rm{K}<T<10^{10}\rm{K}$). The luminosity is dominated by gas at $10^{7-8}$K, therefore its spectral temperature is $\sim 2\times 10^7$ K ($\equiv 2.0$ keV). However, the galaxy hosts significant amount of mass at much lower temperatures. While this lower temperature gas ($T<10^6$ K) does not contribute significantly to the total luminosity, it contributes towards lowering \tew. As a result, \tspec\ is greater than \tew.

The right panel of Figure \ref{fig:histo_lx} corresponds to a time step for which \tspec\ predicts \tew\ correctly (both equal to $\sim 1.9$ keV). At this time step, the luminosity distribution in the temperature-density plane is relatively narrow and though there is mass at higher temperatures ($>10^6$ K), it does not dominate the emission. As a result, both the temperatures are close to each other. The reason why there is copious amount of extremely hot at $0.275$ Gyr is because it corresponds to a local peak in the luminosity (perhaps corresponding to a peak in the luminosity). We conclude here that \tspec, estimated using the single temperature spectral fitting model, reliably predicts \tew\ only when the gas in the galaxy possesses a narrow temperature distribution. 

\subsection{Fitting Log Normal Model to Spectra}

\begin{figure}
	\includegraphics[width=\columnwidth]{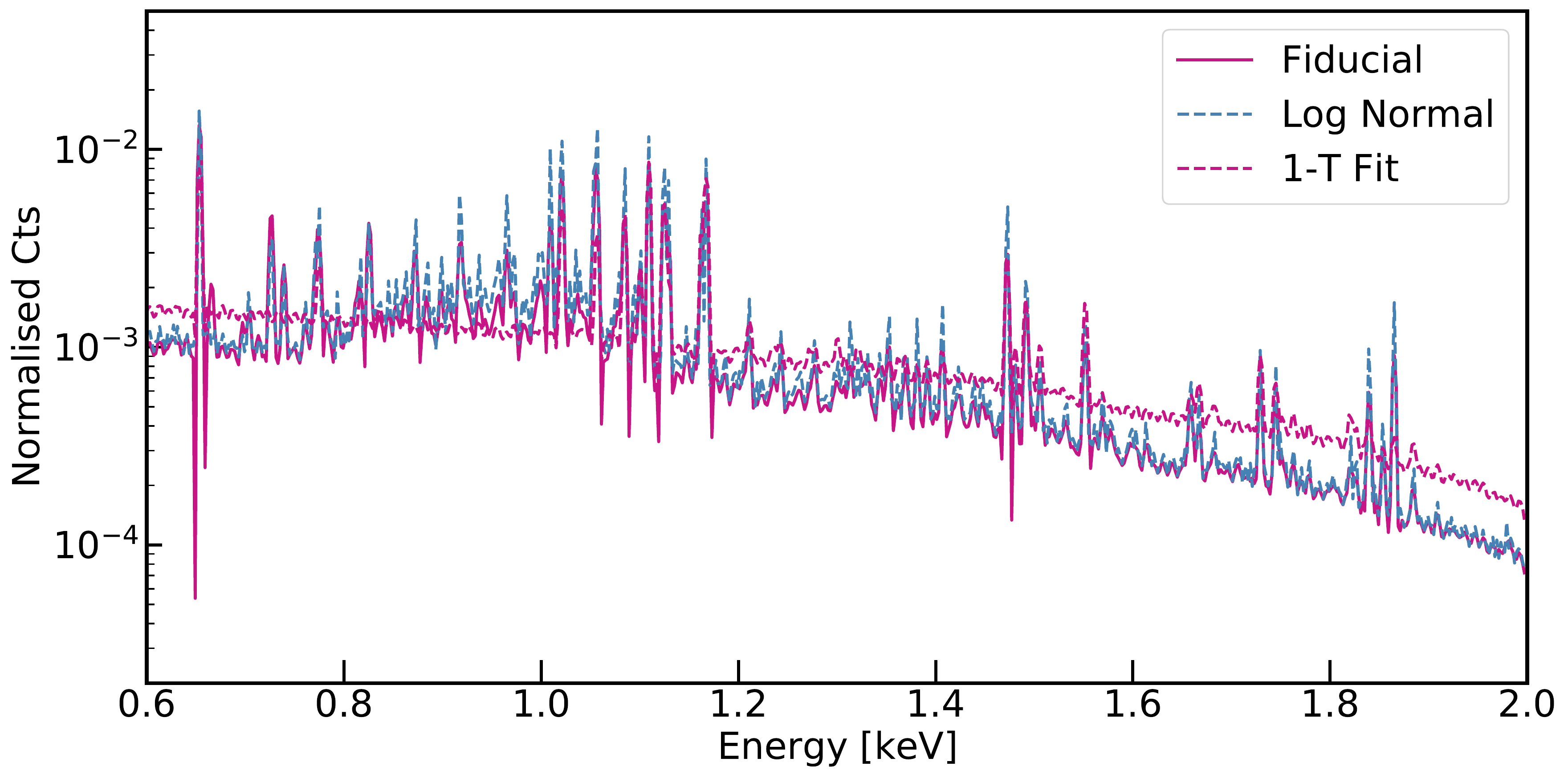}
    \caption{Comparison of the spectrum from the log normal model of temperature distribution, the Fiducial model at the same time step and the fit corresponding to the single temperature fit.}
    \label{fig:log_normal}
\end{figure}

In the Section \ref{sec:compare_tspec_tew}, we show that the discrepancy between \tspec\ and \tew\ is considerable if the actual temperature distribution of the underlying gas is over several orders of magnitude. In such a case, fitting the spectrum with a single temperature emission model is not entirely physical. It has been suggested that a log-normal temperature distribution is better suited to represent gas that occupies a wide range in temperature \citep{Vijayan&Li22}. To assess this, we re-fit the spectrum at the time step, represented by the teal square in Figure \ref{fig:tspec_tew_0-10kpc}, at which there is large discrepancy between \tspec\ and \tew. 

As indicated in the left panel of Figure \ref{fig:histo_lx}, the X-ray emitting gas lies between $\sim 10^{7-10}$ K. To obtain a spectrum from a log-normal distribution, we construct a box having dimensions identical to the simulation domain. The values of the density and metallicity fields are same as that of the simulation box. For every cell in the box, we extract a temperature value from a log-normal number distribution with a peak temperature at $\sim 2\times 10^7$ K and a width of $0.4$ dex in temperature. We choose the peak temperature from mass-weighted temperature distribution of the gas and the width of the log-normal is identical to that used in the analysis of \cite{Vijayan&Li22}. We follow the procedure described in Section \ref{sec:extract_pure_spectra} to extract the spectrum from the log-normal temperature distribution.

In Figure \ref{fig:log_normal}, we show the spectrum from the simulation (solid curve), the single temperature fit as well as the from the log-normal temperature distribution (dashed curves). We note here that for this particular time step, the \tspec\ from single temperature fit is $\sim 3$ keV, while \tew$\sim 0.3$ keV. As can be seen from Figure \ref{fig:log_normal}, the single temperature fit is unable to reproduce the simulation spectrum at both high and low energies. However, the spectrum generated from the log-normal distribution is able to reproduce the spectrum well across the entire energy range because it is able to capture the wide temperature distribution of the gas.

\subsection{Radial Distribution of \tspec}

\begin{figure}
	\includegraphics[width=\columnwidth]{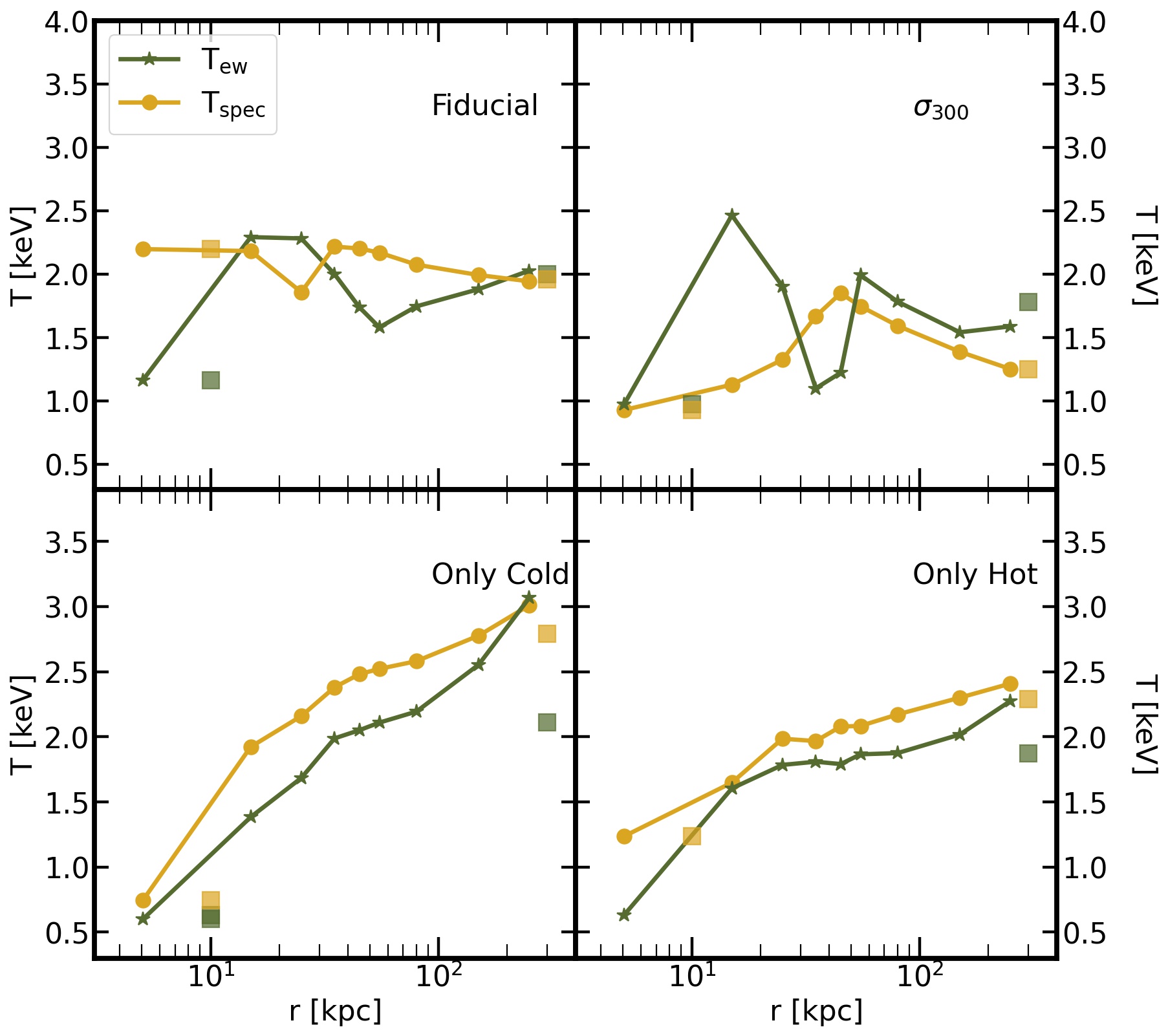}
    \caption{Radial distribution of \tspec\ and \tew\ for all the runs. We produce \tspec\ and \tew\ for radial bins that are $10$ kpc wide. Squares show the \tspec\ and \tew\ values for ISM and CGM, plotted at $10$ and $300$ kpc, respectively. We note that \tspec\ is a better representative of \tew\ in smaller radial bins.}
    \label{fig:tew-tspec-rad}
\end{figure}

To understand if extracting \tspec\ from smaller regions results in better match with \tew, we use the analysis described in the previous Section to generate the radial distribution of \tspec. Now, instead of bisecting the simulation domain into ISM and CGM, we now extract spectra from annular regions which are $10$ kpc wide. We proceed as previously and obtain \tspec\ and \tew\ for these regions. 

In Figure \ref{fig:tew-tspec-rad}, we plot the \tspec\ (golden) and \tew\ (olive) for the annular regions against the mid-point radius for the region. We also show the \tspec\ and \tew\ values for the ISM and CGM regions in square. We note that for all the runs the curves for \tspec\ and \tew\ are within a factor of $2$, even when the aggregate estimates for CGM differ by a larger margin (see bottom left panel for ``Only Cold''). 
For the Fiducial run, \tspec\ and \tew\ match for the CGM, while they differ by a factor of $\sim 2$ in the ISM. We note that the mismatch between the CGM values of \tspec\ and \tew\ is larger than the corresponding values in any radial bin. The reason is because to estimate the CGM \tspec\ and \tew\ values we average over a large volume resulting in averaging over a larger range of temperature values which may not be a representative of temperatures at all radii. 

From this Figure, we can conclude that \tspec\ is a better prediction of \tew\ in smaller radial bins. This is not surprising because smaller spatial volumes for averaging lead to a narrower range over which gas temperature varies, resulting in distributions that closely resemble 1-T distributions. Such an effect has been discussed previously in \cite{Vijayan&Li22}.

\subsection{Mock HUBS Spectrum}

\begin{figure}
	\includegraphics[width=\columnwidth]{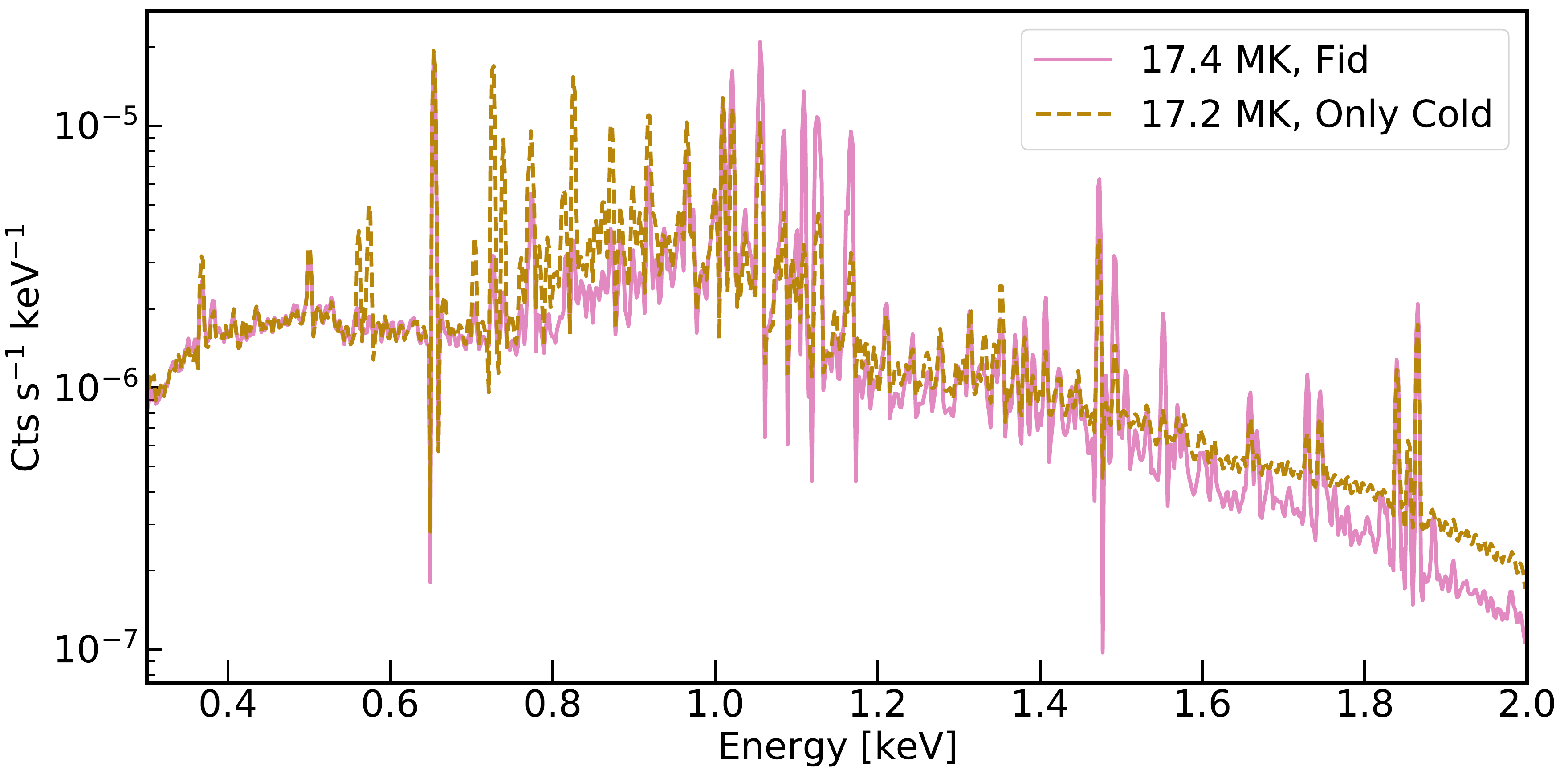}
    \caption{The spectra from the Fiducial and ``Only Cold'' simulations between $0.1$ and $10$ kpc. The spectra have been normalised with respective total luminosity.}
    \label{fig:spec_compare}
\end{figure}

\begin{figure}
	\includegraphics[width=\columnwidth]{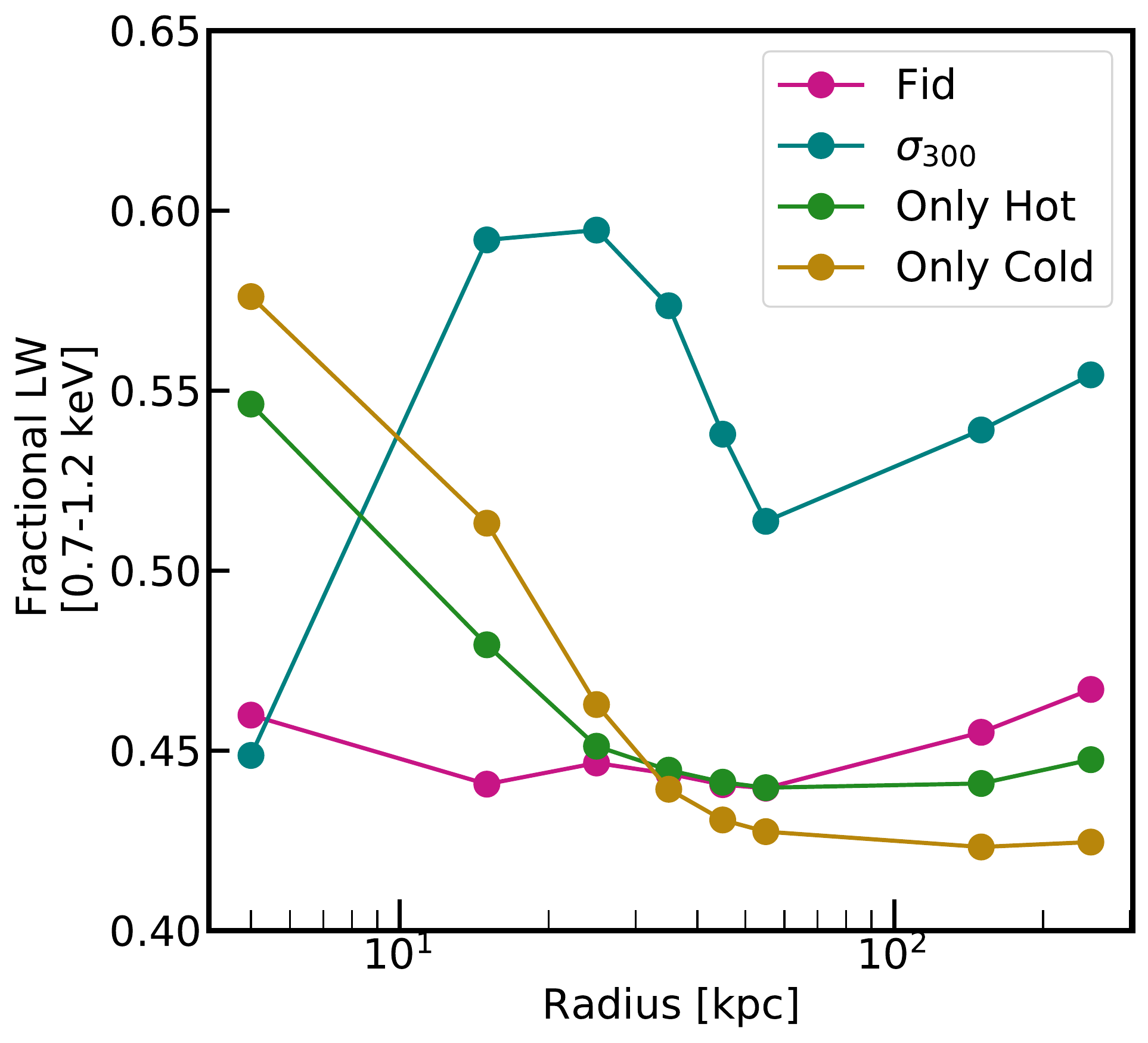}
    \caption{The ratio of the counts between $0.7-1.2$ keV and the total counts for $0.2-2.0$ keV for different runs. }
    \label{fig:frac_lw}
\end{figure}

In Figure \ref{fig:spec_compare}, we show the high-resolution spectrum for the Fiducial and ``Only Cold'' models at the same time step and fit by nearly identical single temperature fits (\tspec$\sim 17$ MK). Despite identical fitting temperature, the spectra are not identical. We note the differences in the $0.7-1.0$ and $\gtrsim 1.6$ keV range. The feature in the $0.7-1.0$ keV range are associated with Fe-line complexes which are very sensitive to the gas temperature \citep{Bohringer&Werner10}. To explore whether such features in the spectrum can be used as a diagnostic for discriminating between AGN models and galaxy sizes, we estimate the fractional $0.7-1.2$ keV line width which is the ratio between the counts in $0.7-1.2$ keV to the counts in the full energy range ($0.3-2.0$ keV) for all the models. To understand how this quantity changes with radius, in Figure \ref{fig:frac_lw} we show it as a function of radius. The binning in radius is narrower ($10$ kpc wide) for $r<60$ kpc and wider ($100$ kpc) for radii larger than $60$ kpc. This is because that temperature and density profiles are have lesser variation for larger radii (Figures \ref{fig:tew_vs_rad} and \ref{fig:nH_vs_rad}). The line-width ratios have been averaged over all the timesteps.

The fractional line-width for \sigth\ shows a trend vastly different from the other three models. Unlike the other three models, at radii close to the centre, the fractional line-width is relatively small. It peaks at around $\sim 20$ kpc, falls off upto $60$ kpc and increase thereafter. The ``Only Hot'' and ``Only Cold'' models show trends similar to each other. The fractional line-width is strongest close to the centre, which has the highest density and temperature values and it flattens out at larger radii. The Fiducial fractional line-width remains nearly flat for the entire radial domain. From this Figure, we can conclude that fractional line-width should be able to distinguish the different feedback models active within the galaxy from the fractional line-widths from the inner $10-20$ kpc.

\subsection{Simulating HUBS Emission}\label{sec:Mock_Image}
\begin{figure}
	\includegraphics[width=\columnwidth]{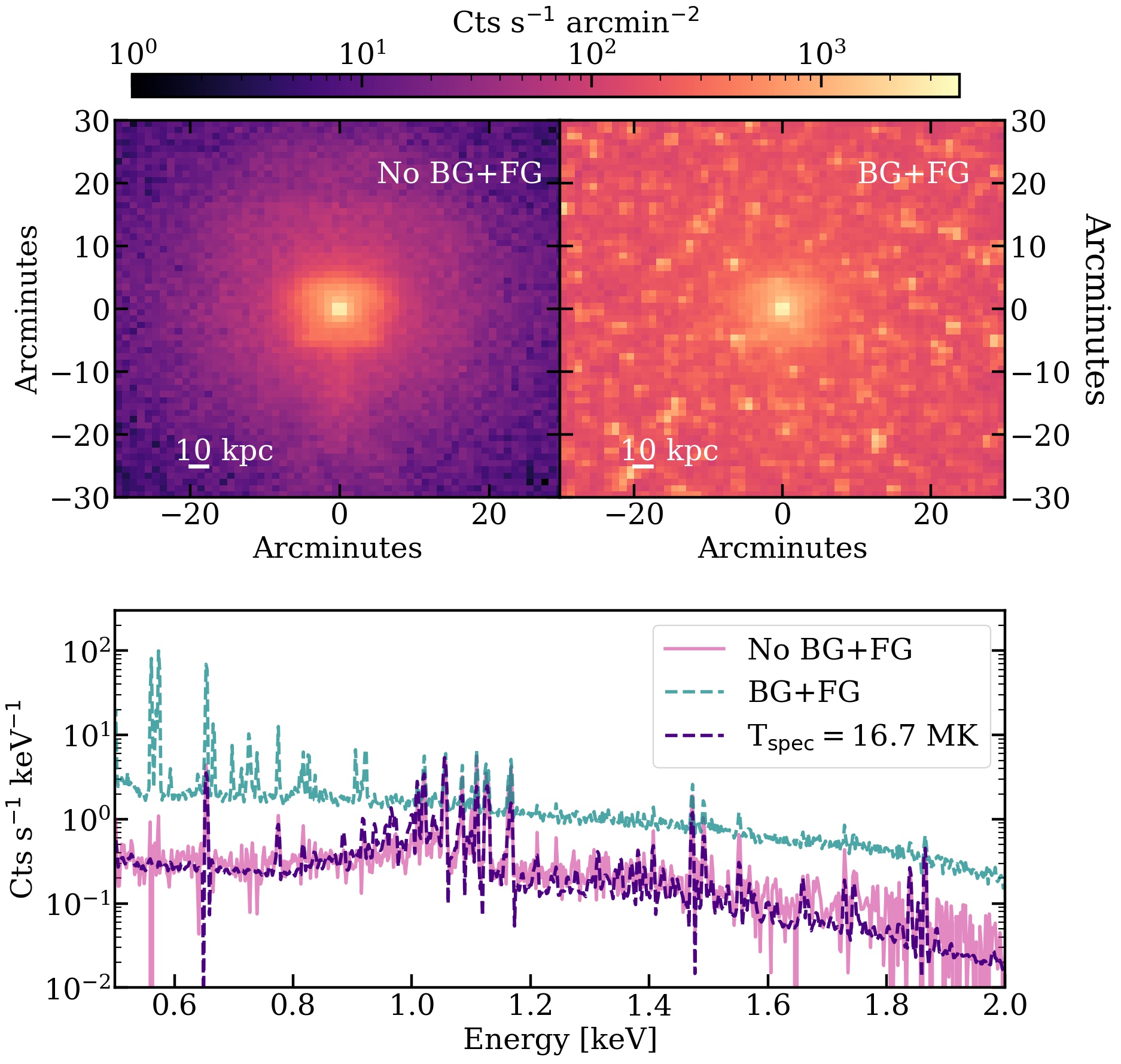}
    \caption{In the top panel, we show mock image of the Fiducial run using HUBS with (left) and without (right) instrument and point source background and foregrounds. This is a $300$ ks observation. The galaxy is at a distance of $15$ Mpc. The bottom panel shows the spectra extracted from the mock image, with and without the background and foreground contributions. The dashed navy curve shows the 1-T fit to the mock galaxy spectrum.}
    \label{fig:mock_image}
\end{figure}

\begin{figure}
	\includegraphics[width=\columnwidth]{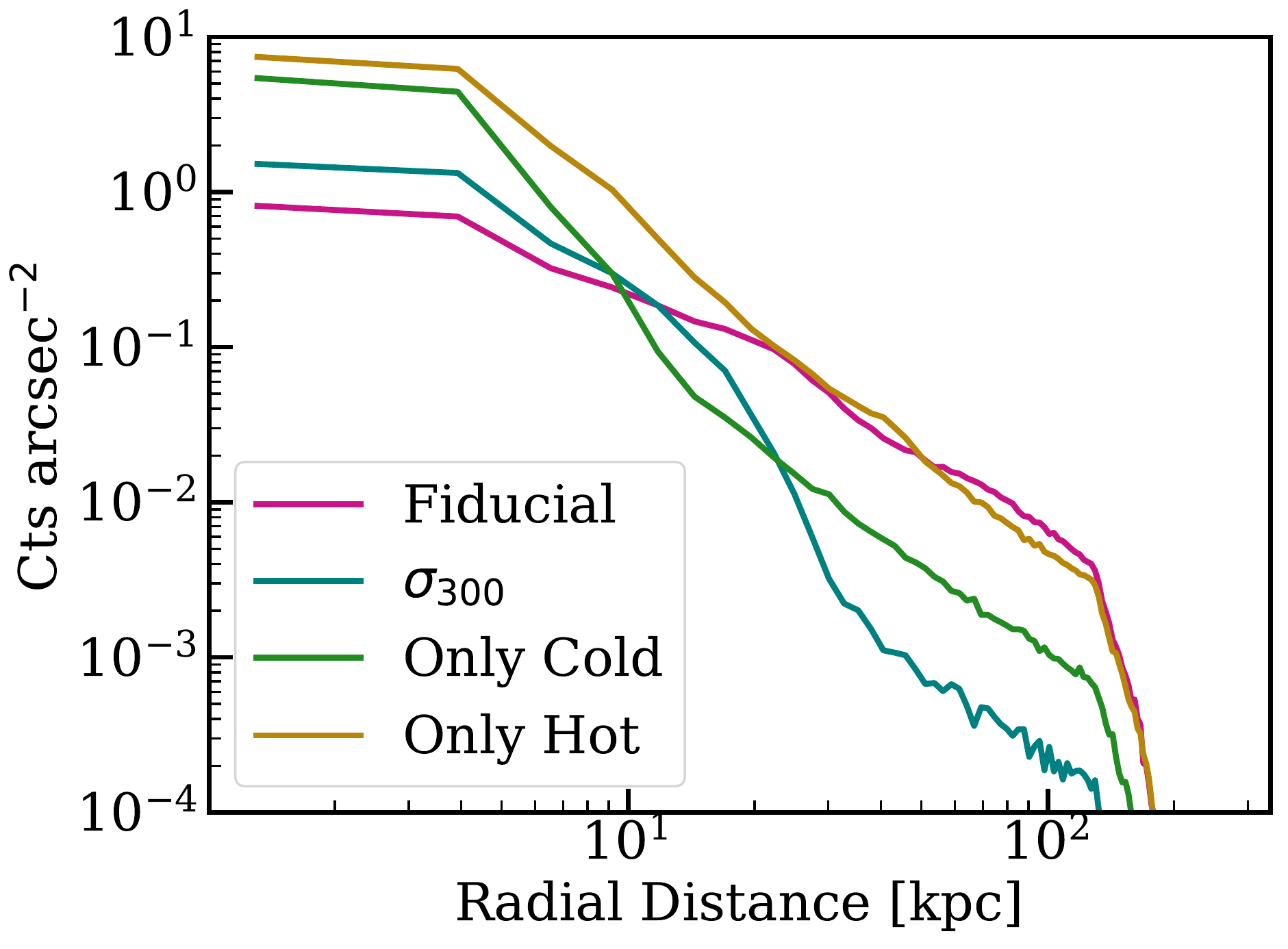}
    \caption{Surface brightness profile of the mock image for the various runs, corresponding to the time step half-way through the simulation at $\sim 1$ Gyr. }
    \label{fig:rad_sb}
\end{figure}

We convert the 2D spherical data from the simulations into 3D Cartesian data using the assumption of axi-symmetry. This conversion is essentially to process the data for an X-ray emission map.
Because this 2D to 3D is a computationally expensive process, we use a much lower resolution for the Cartesian data. 

We use the python package, pyXSIM\footnote{\url{https://hea-www.cfa.harvard.edu/~jzuhone/pyxsim/}}, for generating photon lists from the simulation datasets. To produce a realistic image, convolved with an instrument response, we rely on SOXS\footnote{\url{https://hea-www.cfa.harvard.edu/soxs/index.html}}. For producing the photon list from pyXSIM we use a distance of $15$ Mpc and an exposure time of $300$ ks. We generate a mock HUBS image of the galaxy, using the Fiducial model. We excise the innermost $600$ pc of the galaxy as it is excessively bright, and produce X-ray emission from the gas within $300$ kpc of the galaxy. We provide arbitrary RA and Dec values. We select a time step ($1.25$ Gyr, which is half way through the simulation) and generate a mock image based on the underlying density and temperature distribution at this time step. The resulting image generated from such a synthetic observation is shown in the top-left panel of Figure \ref{fig:mock_image}. We note here that we do not include any contribution from various backgrounds and foregrounds for this image. As expected, the region close to the centre of the galaxy is the brightest and the brightness decreases radially outwards. This is expected since most of the volume of the outer-CGM is filled with low density gas (top-right panels of Figures \ref{fig:nH_vs_rad} and \ref{fig:tew_vs_rad}). We therefore expect the X-ray brightness to fall off sharply away from the centre. At the scale shown in the top panel of Figure \ref{fig:mock_image}, the distinguishing features in the gas density and temperature profiles are smoothed out.

SOXS allows users to turn on contribution from foreground and the instrument and point source background\footnote{We discuss the effect of including foreground and background  to the signal from the galaxy in detail in Appendix \ref{sec:fgbk_appendix}. The details on the foreground and background models used in SOXS can be found at \url{https://hea-www.cfa.harvard.edu/soxs/users_guide/background.html}.}. In the top-right panel of Figure \ref{fig:mock_image}, we show the mock image produced after taking into account foreground and background contributions. HUBS has a large FoV resulting in a background emission that is stronger than the source by nearly two orders \citep{Vijayan&Li22}. A similar conclusion can be drawn from the left panel of Figure \ref{fig:mock_image}. With the inclusion of foreground and background, the extended features of the X-ray emission are lost and only the innermost $\sim 20$ kpc are distinguishable. The signal from the galaxies could be redshifted with respect to the foreground at $z=0$ and thus be differentiated.  

We process the event files, generated for the mock images, to extract the spectra which we show in the bottom panel of Figure \ref{fig:mock_image}. We show this spectrum in the teal curve in the bottom panel of Figure \ref{fig:mock_image}.  To extract the galaxy spectrum, we need to remove the contribution to the total spectrum from the background and foreground. We use SOXS to generate an event file for an observation with the contribution from galaxy turned off and hence obtain the background+foreground spectrum. We then subtract this from the full mock spectrum to get the mock galaxy spectrum. We fit this mock galaxy spectrum using the procedure described in Section \ref{sec:extract_pure_spectra}. The fitting temperature, assuming a 1-T emission model, is $\sim 16.7$ MK. The fitted spectrum is plotted in the dashed navy, while the pure galaxy signal is depicted in the magenta curve. As we discuss in Appendix \ref{sec:fgbk_appendix}, the background and foreground emission is much stronger than that from the source. Overall, we find that the 1-T model provides a reasonable fit to the mock spectrum. It should be noted here that there are discrepancies in the strengths of certain individual lines, especially around $\sim 1$ keV. We note here that for the timestep in the simulation used in this analysis \tew$\sim 22$ MK.  

We repeat the process and produce mock images for all the models discussed in this paper (without foreground and background emission). For each of these runs, we radially average the surface brightness profile and obtain a radial intensity plot. We divide the pixels of the mock image into equal radially spaced $100$ annular regions and sum up the total number of photon counts in each region. We divide the total photon count by the area of the annular region to normalise and plot it against the physical distance along the radius of the galaxy. We show the annulus-averaged brightness in Figure \ref{fig:rad_sb}. Though \tspec\ values for ``Only Cold'' and ``Only Hot'' models are similar, $18$ and $14$ MK, respectively, their radial profiles have different slopes, indicating that radial surface brightness profiles might hold clue to the mode of accretion taking place in the galaxy's AGN.

\section{Discussion \& Conclusions}\label{sec:discussion_and_conclusion}

\subsection{Comparison with Other Works}

We have undertaken a systematic study of the relationship between two estimates of gas temperatures, viz, \tspec\ and \tew. While \tspec\ is obtained from spectral analysis of X-ray emission from hot gas in the ISM and the CGM around elliptical galaxies, \tew\ is estimated from the simulation data. The properties of the hot gas are intrinsically related to the properties of the SMBH hosted by the galaxy and the correlations between the two have been studied via observations \citep{Lakhchaura+18, Gaspari+19, Lakhchaura+19} and simulations \citep{Gaspari+14,Truong+20, Truong+20A}. While these works focus on quantities such as the black hole mass, the total X-ray luminosity etc, our work aims to understand the spectral properties of the emission. Such a study is critical in the context of the upcoming X-ray telescope, HUBS, which has a large FoV ($1^{\circ}\times 1^{\circ}$ and $\sim$ eV spectral resolution (compared to $\sim 130$ eV of \textit{Chandra}). 

The high resolution of the X-ray spectrum will provide us with abundant information about the temperature distribution of the hot gas. In this respect, it is critical to evaluate the relationships between the extracted spectral temperature and the actual gas temperature. \cite{Mazzotta+04} address the discrepancy between the \textit{Chandra} spectral temperatures of galaxy clusters and their physical equivalents from simulations. They find that the spectroscopic temperature obtained from X-ray observations is always lower than the emission-weighted temperature in the cluster. In our analysis, we find that \tspec\ may be higher than \tew\ by a factor of $\sim 4$ (Figures \ref{fig:tspec_tew_0-10kpc} and \ref{fig:tspec_tew_10-300kpc}).


\subsection{Conclusions}

We have analysed 2D axi-symmertic simulations of the evolution of an elliptical galaxy under the influence of feedback from the SMBH at its centre. We are interested in the X-ray emission from the hot ($T>10^6$ K) diffuse gas in the ISM and the CGM of such a galaxy. We explore four sets of simulations representing the different forms of feedback (``Fiducial'', ``Only Hot'', and ``Only Cold'') and different galaxy size (``\sigth\ ''). We follow the galaxy evolution for a period of $\sim 2$ Gyr, over which the SMBH undergoes several outburst phases resulting in radially declining density and temperature profiles (Figures \ref{fig:nH_vs_rad} and \ref{fig:tew_vs_rad}). Because of the stochasticity of the outbursts, the soft X-ray luminosity ($0.3-5.0$ keV) varies considerably over the simulation time period (Figure \ref{fig:lx_vs_t}. We use pyAtomDB for estimating the spectral temperature of the gas in the simulation for low and high spectral resolutions and compare it with the emission weighted temperature. 

Our main conclusions are as follows-
\begin{enumerate}
    
    \item The spectral temperature, estimated using spectral analysis, is different from the emission-weighted temperature by a factor of few.
    
    \item The low- (resolution $\sim 130$ eV) and high-resolution (resolution $\sim 2$ eV) instruments produce nearly the same  predictions for the \tspec.
    
    \item The difference between \tspec\ and \tew\ arise because the spectral fitting model (a single temperature fit) is not able to capture the gas distribution accurately. Using a more physically motivated model, such as the log-normal model, can potentially alleviate the discrepancies between the two temperatures (Figure \ref{fig:log_normal}).
    
    \item Even if \tspec\ is similar for different models, the underlying gas properties might be different. Such differences appear only upon analysis of the full spectra (Figure \ref{fig:spec_compare}).
    
    \item The ratio of counts between $0.7-1.2$ keV, a range corresponding to Fe-line emission, can potentially be a diagnostic tool for discriminating between different accretion models (Figure \ref{fig:frac_lw}).
    
    \item The surface brightness maps could also hold clues about the exact mode of accretion taking place within the galaxy, as the radial profiles of surface brightness possess different slopes for the various runs (Figure \ref{fig:rad_sb}).
\end{enumerate}

For future studies, we plan to expand this work by focussing on a wide range of galaxy masses. Given that we expect high resolution spectral data from HUBS, 1-T and log-normal fits are rather simplistic. A much more detailed analysis can be pursued once spectral fitting models are developed for HUBS.  

\section*{Acknowledgements}
We thank Drs. Wei Cui and Jiangtao Li for helpful discussions and comments. We also thank the referee for their useful comments and suggestions. AV, BZ, and FY are supported in part by the Natural Science Foundation of China (grants 12133008, 12192220, and 12192223), and the China Manned Space Project (No. CMS-CSST-2021-B02). LCH was supported by the National Science Foundation of China (11721303, 11991052, 12011540375) and the China Manned Space Project (CMS-CSST-2021-A04, CMS-CSST-2021-A06). The analysis, presented in this paper, was done using the High Performance Computing Resource in the Core Facility for Advanced Research Computing
at Shanghai Astronomical Observatory. AV would like to thank the SHAO staff maintaining the facility for their support.

\section*{Data Availability}
The data underlying this paper will be shared on reasonable request to the corresponding author.






\begin{thebibliography}{}
\makeatletter
\relax
\def\mn@urlcharsother{\let\do\@makeother \do\$\do\&\do\#\do\^\do\_\do\%\do\~}
\def\mn@doi{\begingroup\mn@urlcharsother \@ifnextchar [ {\mn@doi@}
  {\mn@doi@[]}}
\def\mn@doi@[#1]#2{\def\@tempa{#1}\ifx\@tempa\@empty \href
  {http://dx.doi.org/#2} {doi:#2}\else \href {http://dx.doi.org/#2} {#1}\fi
  \endgroup}
\def\mn@eprint#1#2{\mn@eprint@#1:#2::\@nil}
\def\mn@eprint@arXiv#1{\href {http://arxiv.org/abs/#1} {{\tt arXiv:#1}}}
\def\mn@eprint@dblp#1{\href {http://dblp.uni-trier.de/rec/bibtex/#1.xml}
  {dblp:#1}}
\def\mn@eprint@#1:#2:#3:#4\@nil{\def\@tempa {#1}\def\@tempb {#2}\def\@tempc
  {#3}\ifx \@tempc \@empty \let \@tempc \@tempb \let \@tempb \@tempa \fi \ifx
  \@tempb \@empty \def\@tempb {arXiv}\fi \@ifundefined
  {mn@eprint@\@tempb}{\@tempb:\@tempc}{\expandafter \expandafter \csname
  mn@eprint@\@tempb\endcsname \expandafter{\@tempc}}}

\bibitem[\protect\citeauthoryear{{Anderson} \& {Bregman}}{{Anderson} \&
  {Bregman}}{2011}]{Anderson+11}
{Anderson} M.~E.,  {Bregman} J.~N.,  2011, \mn@doi [\apj]
  {10.1088/0004-637X/737/1/22}, \href
  {https://ui.adsabs.harvard.edu/abs/2011ApJ...737...22A} {737, 22}

\bibitem[\protect\citeauthoryear{{Anderson}, {Gaspari}, {White}, {Wang}  \&
  {Dai}}{{Anderson} et~al.}{2015}]{Anderson+15}
{Anderson} M.~E.,  {Gaspari} M.,  {White} S. D.~M.,  {Wang} W.,   {Dai} X.,
  2015, \mn@doi [\mnras] {10.1093/mnras/stv437}, \href
  {https://ui.adsabs.harvard.edu/abs/2015MNRAS.449.3806A} {449, 3806}

\bibitem[\protect\citeauthoryear{{Anderson}, {Churazov}  \&
  {Bregman}}{{Anderson} et~al.}{2016}]{Anderson+16}
{Anderson} M.~E.,  {Churazov} E.,   {Bregman} J.~N.,  2016, \mn@doi [\mnras]
  {10.1093/mnras/stv2314}, \href
  {https://ui.adsabs.harvard.edu/abs/2016MNRAS.455..227A} {455, 227}

\bibitem[\protect\citeauthoryear{{Babyk}, {McNamara}, {Nulsen}, {Hogan},
  {Vantyghem}, {Russell}, {Pulido}  \& {Edge}}{{Babyk} et~al.}{2018}]{Babyk+18}
{Babyk} I.~V.,  {McNamara} B.~R.,  {Nulsen} P.~E.~J.,  {Hogan} M.~T.,
  {Vantyghem} A.~N.,  {Russell} H.~R.,  {Pulido} F.~A.,   {Edge} A.~C.,  2018,
  \mn@doi [\apj] {10.3847/1538-4357/aab3c9}, \href
  {https://ui.adsabs.harvard.edu/abs/2018ApJ...857...32B} {857, 32}

\bibitem[\protect\citeauthoryear{{Bogd{\'a}n}, {Forman}, {Kraft}  \&
  {Jones}}{{Bogd{\'a}n} et~al.}{2013a}]{Bogdan+13}
{Bogd{\'a}n} {\'A}.,  {Forman} W.~R.,  {Kraft} R.~P.,   {Jones} C.,  2013a,
  \mn@doi [\apj] {10.1088/0004-637X/772/2/98}, \href
  {https://ui.adsabs.harvard.edu/abs/2013ApJ...772...98B} {772, 98}

\bibitem[\protect\citeauthoryear{{Bogd{\'a}n}, {Forman}, {Kraft}  \&
  {Jones}}{{Bogd{\'a}n} et~al.}{2013b}]{Bogdan+13B}
{Bogd{\'a}n} {\'A}.,  {Forman} W.~R.,  {Kraft} R.~P.,   {Jones} C.,  2013b,
  \mn@doi [\apj] {10.1088/0004-637X/772/2/98}, \href
  {https://ui.adsabs.harvard.edu/abs/2013ApJ...772...98B} {772, 98}

\bibitem[\protect\citeauthoryear{{Bogd{\'a}n}, {Bourdin}, {Forman}, {Kraft},
  {Vogelsberger}, {Hernquist}  \& {Springel}}{{Bogd{\'a}n}
  et~al.}{2017}]{Bogdan+17}
{Bogd{\'a}n} {\'A}.,  {Bourdin} H.,  {Forman} W.~R.,  {Kraft} R.~P.,
  {Vogelsberger} M.,  {Hernquist} L.,   {Springel} V.,  2017, \mn@doi [\apj]
  {10.3847/1538-4357/aa9523}, \href
  {https://ui.adsabs.harvard.edu/abs/2017ApJ...850...98B} {850, 98}

\bibitem[\protect\citeauthoryear{{Bogd{\'a}n}, {Lovisari}, {Volonteri}  \&
  {Dubois}}{{Bogd{\'a}n} et~al.}{2018}]{Bogdan+18}
{Bogd{\'a}n} {\'A}.,  {Lovisari} L.,  {Volonteri} M.,   {Dubois} Y.,  2018,
  \mn@doi [\apj] {10.3847/1538-4357/aa9ab5}, \href
  {https://ui.adsabs.harvard.edu/abs/2018ApJ...852..131B} {852, 131}

\bibitem[\protect\citeauthoryear{{B{\"o}hringer} \& {Werner}}{{B{\"o}hringer}
  \& {Werner}}{2010}]{Bohringer&Werner10}
{B{\"o}hringer} H.,  {Werner} N.,  2010, \mn@doi [\aapr]
  {10.1007/s00159-009-0023-3}, \href
  {https://ui.adsabs.harvard.edu/abs/2010A&ARv..18..127B} {18, 127}

\bibitem[\protect\citeauthoryear{{Booth} \& {Schaye}}{{Booth} \&
  {Schaye}}{2011}]{Booth&Schaye11}
{Booth} C.~M.,  {Schaye} J.,  2011, \mn@doi [\mnras]
  {10.1111/j.1365-2966.2011.18203.x}, \href
  {https://ui.adsabs.harvard.edu/abs/2011MNRAS.413.1158B} {413, 1158}

\bibitem[\protect\citeauthoryear{{Boroson}, {Kim}  \& {Fabbiano}}{{Boroson}
  et~al.}{2011}]{Boroson+11}
{Boroson} B.,  {Kim} D.-W.,   {Fabbiano} G.,  2011, \mn@doi [\apj]
  {10.1088/0004-637X/729/1/12}, \href
  {https://ui.adsabs.harvard.edu/abs/2011ApJ...729...12B} {729, 12}

\bibitem[\protect\citeauthoryear{{Cheung} et~al.,}{{Cheung}
  et~al.}{2016}]{Cheung2016}
{Cheung} E.,  et~al., 2016, \mn@doi [\nat] {10.1038/nature18006}, \href
  {https://ui.adsabs.harvard.edu/abs/2016Natur.533..504C} {533, 504}

\bibitem[\protect\citeauthoryear{{Choi}, {Ostriker}, {Naab}, {Oser}  \&
  {Moster}}{{Choi} et~al.}{2015}]{Choi+15}
{Choi} E.,  {Ostriker} J.~P.,  {Naab} T.,  {Oser} L.,   {Moster} B.~P.,  2015,
  \mn@doi [\mnras] {10.1093/mnras/stv575}, \href
  {https://ui.adsabs.harvard.edu/abs/2015MNRAS.449.4105C} {449, 4105}

\bibitem[\protect\citeauthoryear{{Cui} et~al.,}{{Cui} et~al.}{2020}]{Cui+20}
{Cui} W.,  et~al., 2020, \mn@doi [Journal of Low Temperature Physics]
  {10.1007/s10909-019-02279-3}, \href
  {https://ui.adsabs.harvard.edu/abs/2020JLTP..199..502C} {199, 502}

\bibitem[\protect\citeauthoryear{{Dai}, {Anderson}, {Bregman}  \&
  {Miller}}{{Dai} et~al.}{2012}]{Dai+12}
{Dai} X.,  {Anderson} M.~E.,  {Bregman} J.~N.,   {Miller} J.~M.,  2012, \mn@doi
  [\apj] {10.1088/0004-637X/755/2/107}, \href
  {https://ui.adsabs.harvard.edu/abs/2012ApJ...755..107D} {755, 107}

\bibitem[\protect\citeauthoryear{{Davies}, {Crain}, {McCarthy}, {Oppenheimer},
  {Schaye}, {Schaller}  \& {McAlpine}}{{Davies} et~al.}{2019}]{Davies+19}
{Davies} J.~J.,  {Crain} R.~A.,  {McCarthy} I.~G.,  {Oppenheimer} B.~D.,
  {Schaye} J.,  {Schaller} M.,   {McAlpine} S.,  2019, \mn@doi [\mnras]
  {10.1093/mnras/stz635}, \href
  {https://ui.adsabs.harvard.edu/abs/2019MNRAS.485.3783D} {485, 3783}

\bibitem[\protect\citeauthoryear{{Fabian}}{{Fabian}}{2012}]{Fabian2012}
{Fabian} A.~C.,  2012, \mn@doi [\araa] {10.1146/annurev-astro-081811-125521},
  \href {https://ui.adsabs.harvard.edu/abs/2012ARA&A..50..455F} {50, 455}

\bibitem[\protect\citeauthoryear{{Foster} \& {Heuer}}{{Foster} \&
  {Heuer}}{2020}]{Foster+20}
{Foster} A.~R.,  {Heuer} K.,  2020, \mn@doi [Atoms] {10.3390/atoms8030049},
  \href {https://ui.adsabs.harvard.edu/abs/2020Atoms...8...49F} {8, 49}

\bibitem[\protect\citeauthoryear{{Gaspari}, {Brighenti}  \& {Temi}}{{Gaspari}
  et~al.}{2012}]{Gaspari+12}
{Gaspari} M.,  {Brighenti} F.,   {Temi} P.,  2012, \mn@doi [\mnras]
  {10.1111/j.1365-2966.2012.21183.x}, \href
  {https://ui.adsabs.harvard.edu/abs/2012MNRAS.424..190G} {424, 190}

\bibitem[\protect\citeauthoryear{{Gaspari}, {Brighenti}, {Temi}  \&
  {Ettori}}{{Gaspari} et~al.}{2014}]{Gaspari+14}
{Gaspari} M.,  {Brighenti} F.,  {Temi} P.,   {Ettori} S.,  2014, \mn@doi
  [\apjl] {10.1088/2041-8205/783/1/L10}, \href
  {https://ui.adsabs.harvard.edu/abs/2014ApJ...783L..10G} {783, L10}

\bibitem[\protect\citeauthoryear{{Gaspari} et~al.,}{{Gaspari}
  et~al.}{2019}]{Gaspari+19}
{Gaspari} M.,  et~al., 2019, \mn@doi [\apj] {10.3847/1538-4357/ab3c5d}, \href
  {https://ui.adsabs.harvard.edu/abs/2019ApJ...884..169G} {884, 169}

\bibitem[\protect\citeauthoryear{{Gofford}, {Reeves}, {McLaughlin}, {Braito},
  {Turner}, {Tombesi}  \& {Cappi}}{{Gofford}
  et~al.}{2015}]{2015MNRAS.451.4169G}
{Gofford} J.,  {Reeves} J.~N.,  {McLaughlin} D.~E.,  {Braito} V.,  {Turner}
  T.~J.,  {Tombesi} F.,   {Cappi} M.,  2015, \mn@doi [\mnras]
  {10.1093/mnras/stv1207}, \href
  {https://ui.adsabs.harvard.edu/abs/2015MNRAS.451.4169G} {451, 4169}

\bibitem[\protect\citeauthoryear{{Goulding} et~al.,}{{Goulding}
  et~al.}{2016}]{Goulding+16}
{Goulding} A.~D.,  et~al., 2016, \mn@doi [\apj] {10.3847/0004-637X/826/2/167},
  \href {https://ui.adsabs.harvard.edu/abs/2016ApJ...826..167G} {826, 167}

\bibitem[\protect\citeauthoryear{{Heuer}, {Foster}  \& {Smith}}{{Heuer}
  et~al.}{2021}]{Heuer+21}
{Heuer} K.,  {Foster} A.~R.,   {Smith} R.,  2021, \mn@doi [\apj]
  {10.3847/1538-4357/abcaff}, \href
  {https://ui.adsabs.harvard.edu/abs/2021ApJ...908....3H} {908, 3}

\bibitem[\protect\citeauthoryear{{Kim} \& {Fabbiano}}{{Kim} \&
  {Fabbiano}}{2013}]{Kim&Fabbiano13}
{Kim} D.-W.,  {Fabbiano} G.,  2013, \mn@doi [\apj]
  {10.1088/0004-637X/776/2/116}, \href
  {https://ui.adsabs.harvard.edu/abs/2013ApJ...776..116K} {776, 116}

\bibitem[\protect\citeauthoryear{{Kim} \& {Fabbiano}}{{Kim} \&
  {Fabbiano}}{2015}]{Kim&Fabbiano15}
{Kim} D.-W.,  {Fabbiano} G.,  2015, \mn@doi [\apj]
  {10.1088/0004-637X/812/2/127}, \href
  {https://ui.adsabs.harvard.edu/abs/2015ApJ...812..127K} {812, 127}

\bibitem[\protect\citeauthoryear{{Lakhchaura} et~al.,}{{Lakhchaura}
  et~al.}{2018}]{Lakhchaura+18}
{Lakhchaura} K.,  et~al., 2018, \mn@doi [\mnras] {10.1093/mnras/sty2565}, \href
  {https://ui.adsabs.harvard.edu/abs/2018MNRAS.481.4472L} {481, 4472}

\bibitem[\protect\citeauthoryear{{Lakhchaura}, {Truong}  \&
  {Werner}}{{Lakhchaura} et~al.}{2019}]{Lakhchaura+19}
{Lakhchaura} K.,  {Truong} N.,   {Werner} N.,  2019, \mn@doi [\mnras]
  {10.1093/mnrasl/slz114}, \href
  {https://ui.adsabs.harvard.edu/abs/2019MNRAS.488L.134L} {488, L134}

\bibitem[\protect\citeauthoryear{{Lehmer} et~al.,}{{Lehmer}
  et~al.}{2012}]{Lehmer+2012}
{Lehmer} B.~D.,  et~al., 2012, \mn@doi [\apj] {10.1088/0004-637X/752/1/46},
  \href {https://ui.adsabs.harvard.edu/abs/2012ApJ...752...46L} {752, 46}

\bibitem[\protect\citeauthoryear{{Li} \& {Wang}}{{Li} \&
  {Wang}}{2013}]{Li&Wang13}
{Li} J.-T.,  {Wang} Q.~D.,  2013, \mn@doi [\mnras] {10.1093/mnras/sts183},
  \href {https://ui.adsabs.harvard.edu/abs/2013MNRAS.428.2085L} {428, 2085}

\bibitem[\protect\citeauthoryear{{Lopez}, {Mathur}, {Nguyen}, {Thompson}  \&
  {Olivier}}{{Lopez} et~al.}{2020}]{Lopez+20}
{Lopez} L.~A.,  {Mathur} S.,  {Nguyen} D.~D.,  {Thompson} T.~A.,   {Olivier}
  G.~M.,  2020, \mn@doi [\apj] {10.3847/1538-4357/abc010}, \href
  {https://ui.adsabs.harvard.edu/abs/2020ApJ...904..152L} {904, 152}

\bibitem[\protect\citeauthoryear{{Ma}, {Roberts}, {Li}  \& {Wang}}{{Ma}
  et~al.}{2019}]{Ma2019}
{Ma} R.-Y.,  {Roberts} S.~R.,  {Li} Y.-P.,   {Wang} Q.~D.,  2019, \mn@doi
  [\mnras] {10.1093/mnras/sty3039}, \href
  {https://ui.adsabs.harvard.edu/abs/2019MNRAS.483.5614M} {483, 5614}

\bibitem[\protect\citeauthoryear{{Mazzotta}, {Rasia}, {Moscardini}  \&
  {Tormen}}{{Mazzotta} et~al.}{2004}]{Mazzotta+04}
{Mazzotta} P.,  {Rasia} E.,  {Moscardini} L.,   {Tormen} G.,  2004, \mn@doi
  [\mnras] {10.1111/j.1365-2966.2004.08167.x}, \href
  {https://ui.adsabs.harvard.edu/abs/2004MNRAS.354...10M} {354, 10}

\bibitem[\protect\citeauthoryear{{McCammon} et~al.,}{{McCammon}
  et~al.}{2002}]{McCammon+2002}
{McCammon} D.,  et~al., 2002, \mn@doi [\apj] {10.1086/341727}, \href
  {https://ui.adsabs.harvard.edu/abs/2002ApJ...576..188M} {576, 188}

\bibitem[\protect\citeauthoryear{{Morganti}}{{Morganti}}{2017}]{Morganti2017}
{Morganti} R.,  2017, \mn@doi [Frontiers in Astronomy and Space Sciences]
  {10.3389/fspas.2017.00042}, \href
  {https://ui.adsabs.harvard.edu/abs/2017FrASS...4...42M} {4, 42}

\bibitem[\protect\citeauthoryear{{Naab} \& {Ostriker}}{{Naab} \&
  {Ostriker}}{2017}]{Ostriker2017}
{Naab} T.,  {Ostriker} J.~P.,  2017, \mn@doi [\araa]
  {10.1146/annurev-astro-081913-040019}, \href
  {https://ui.adsabs.harvard.edu/abs/2017ARA&A..55...59N} {55, 59}

\bibitem[\protect\citeauthoryear{{Narayan}, {S{\"A} dowski}, {Penna}  \&
  {Kulkarni}}{{Narayan} et~al.}{2012}]{Narayan2012}
{Narayan} R.,  {S{\"A} dowski} A.,  {Penna} R.~F.,   {Kulkarni} A.~K.,  2012,
  \mn@doi [\mnras] {10.1111/j.1365-2966.2012.22002.x}, \href
  {https://ui.adsabs.harvard.edu/abs/2012MNRAS.426.3241N} {426, 3241}

\bibitem[\protect\citeauthoryear{{Park}, {Hada}, {Kino}, {Nakamura}, {Ro}  \&
  {Trippe}}{{Park} et~al.}{2019}]{2019ApJ...871..257P}
{Park} J.,  {Hada} K.,  {Kino} M.,  {Nakamura} M.,  {Ro} H.,   {Trippe} S.,
  2019, \mn@doi [\apj] {10.3847/1538-4357/aaf9a9}, \href
  {https://ui.adsabs.harvard.edu/abs/2019ApJ...871..257P} {871, 257}

\bibitem[\protect\citeauthoryear{{Rasia}, {Mazzotta}, {Borgani}, {Moscardini},
  {Dolag}, {Tormen}, {Diaferio}  \& {Murante}}{{Rasia} et~al.}{2005}]{Rasia+05}
{Rasia} E.,  {Mazzotta} P.,  {Borgani} S.,  {Moscardini} L.,  {Dolag} K.,
  {Tormen} G.,  {Diaferio} A.,   {Murante} G.,  2005, \mn@doi [\apjl]
  {10.1086/427554}, \href
  {https://ui.adsabs.harvard.edu/abs/2005ApJ...618L...1R} {618, L1}

\bibitem[\protect\citeauthoryear{Rubner, Tomasi  \& Guibas}{Rubner
  et~al.}{1998}]{Rubner+98}
Rubner Y.,  Tomasi C.,   Guibas L.,  1998, in Sixth International Conference on
  Computer Vision (IEEE Cat. No.98CH36271). pp 59--66,
  \mn@doi{10.1109/ICCV.1998.710701}

\bibitem[\protect\citeauthoryear{{Schaye} et~al.,}{{Schaye}
  et~al.}{2015}]{Schaye+15}
{Schaye} J.,  et~al., 2015, \mn@doi [\mnras] {10.1093/mnras/stu2058}, \href
  {https://ui.adsabs.harvard.edu/abs/2015MNRAS.446..521S} {446, 521}

\bibitem[\protect\citeauthoryear{{Shi}, {Li}, {Yuan}  \& {Zhu}}{{Shi}
  et~al.}{2021}]{Shi2021}
{Shi} F.,  {Li} Z.,  {Yuan} F.,   {Zhu} B.,  2021, \mn@doi [Nature Astronomy]
  {10.1038/s41550-021-01394-0}, \href
  {https://ui.adsabs.harvard.edu/abs/2021NatAs...5..928S} {5, 928}

\bibitem[\protect\citeauthoryear{{Shi}, {Zhu}, {Li}  \& {Yuan}}{{Shi}
  et~al.}{2022}]{2022ApJ...926..209S}
{Shi} F.,  {Zhu} B.,  {Li} Z.,   {Yuan} F.,  2022, \mn@doi [\apj]
  {10.3847/1538-4357/ac4789}, \href
  {https://ui.adsabs.harvard.edu/abs/2022ApJ...926..209S} {926, 209}

\bibitem[\protect\citeauthoryear{{Sijacki}, {Springel}, {Di Matteo}  \&
  {Hernquist}}{{Sijacki} et~al.}{2007}]{Sijacki+07}
{Sijacki} D.,  {Springel} V.,  {Di Matteo} T.,   {Hernquist} L.,  2007, \mn@doi
  [\mnras] {10.1111/j.1365-2966.2007.12153.x}, \href
  {https://ui.adsabs.harvard.edu/abs/2007MNRAS.380..877S} {380, 877}

\bibitem[\protect\citeauthoryear{{Teyssier}, {Moore}, {Martizzi}, {Dubois}  \&
  {Mayer}}{{Teyssier} et~al.}{2011}]{Teyssier+11}
{Teyssier} R.,  {Moore} B.,  {Martizzi} D.,  {Dubois} Y.,   {Mayer} L.,  2011,
  \mn@doi [\mnras] {10.1111/j.1365-2966.2011.18399.x}, \href
  {https://ui.adsabs.harvard.edu/abs/2011MNRAS.414..195T} {414, 195}

\bibitem[\protect\citeauthoryear{{Truong}, {Pillepich}  \& {Werner}}{{Truong}
  et~al.}{2020a}]{Truong+20A}
{Truong} N.,  {Pillepich} A.,   {Werner} N.,  2020a, arXiv e-prints, \href
  {https://ui.adsabs.harvard.edu/abs/2020arXiv200906634T} {p. arXiv:2009.06634}

\bibitem[\protect\citeauthoryear{{Truong} et~al.,}{{Truong}
  et~al.}{2020b}]{Truong+20}
{Truong} N.,  et~al., 2020b, \mn@doi [\mnras] {10.1093/mnras/staa685}, \href
  {https://ui.adsabs.harvard.edu/abs/2020MNRAS.494..549T} {494, 549}

\bibitem[\protect\citeauthoryear{{Vijayan} \& {Li}}{{Vijayan} \&
  {Li}}{2022}]{Vijayan&Li22}
{Vijayan} A.,  {Li} M.,  2022, \mn@doi [\mnras] {10.1093/mnras/stab3413}, \href
  {https://ui.adsabs.harvard.edu/abs/2022MNRAS.510..568V} {510, 568}

\bibitem[\protect\citeauthoryear{{Vikhlinin}}{{Vikhlinin}}{2006}]{Vikhlinin06}
{Vikhlinin} A.,  2006, \mn@doi [\apj] {10.1086/500121}, \href
  {https://ui.adsabs.harvard.edu/abs/2006ApJ...640..710V} {640, 710}

\bibitem[\protect\citeauthoryear{{Wang} et~al.,}{{Wang}
  et~al.}{2013}]{Wang2013}
{Wang} Q.~D.,  et~al., 2013, \mn@doi [Science] {10.1126/science.1240755}, \href
  {https://ui.adsabs.harvard.edu/abs/2013Sci...341..981W} {341, 981}

\bibitem[\protect\citeauthoryear{{Weinberger} et~al.,}{{Weinberger}
  et~al.}{2017}]{2017MNRAS.465.3291W}
{Weinberger} R.,  et~al., 2017, \mn@doi [\mnras] {10.1093/mnras/stw2944}, \href
  {https://ui.adsabs.harvard.edu/abs/2017MNRAS.465.3291W} {465, 3291}

\bibitem[\protect\citeauthoryear{{Werner}, {McNamara}, {Churazov}  \&
  {Scannapieco}}{{Werner} et~al.}{2019}]{Werner+19}
{Werner} N.,  {McNamara} B.~R.,  {Churazov} E.,   {Scannapieco} E.,  2019,
  \mn@doi [\ssr] {10.1007/s11214-018-0571-9}, \href
  {https://ui.adsabs.harvard.edu/abs/2019SSRv..215....5W} {215, 5}

\bibitem[\protect\citeauthoryear{{Wu}, {Mo}, {Li}  \& {Lim}}{{Wu}
  et~al.}{2020}]{Wu+20}
{Wu} X.,  {Mo} H.,  {Li} C.,   {Lim} S.,  2020, \mn@doi [\apj]
  {10.3847/1538-4357/abb80d}, \href
  {https://ui.adsabs.harvard.edu/abs/2020ApJ...903...26W} {903, 26}

\bibitem[\protect\citeauthoryear{{Yamasaki}, {Sato}, {Mitsuishi}  \&
  {Ohashi}}{{Yamasaki} et~al.}{2009}]{Yamasaki+09}
{Yamasaki} N.~Y.,  {Sato} K.,  {Mitsuishi} I.,   {Ohashi} T.,  2009, \mn@doi
  [\pasj] {10.1093/pasj/61.sp1.S291}, \href
  {https://ui.adsabs.harvard.edu/abs/2009PASJ...61S.291Y} {61, S291}

\bibitem[\protect\citeauthoryear{{Yang}, {Yuan}, {Yuan}  \& {White}}{{Yang}
  et~al.}{2021}]{Yang2021}
{Yang} H.,  {Yuan} F.,  {Yuan} Y.-F.,   {White} C.~J.,  2021, \mn@doi [\apj]
  {10.3847/1538-4357/abfe63}, \href
  {https://ui.adsabs.harvard.edu/abs/2021ApJ...914..131Y} {914, 131}

\bibitem[\protect\citeauthoryear{{Yao}, {Yuan}  \& {Ostriker}}{{Yao}
  et~al.}{2021}]{Yao+21}
{Yao} Z.,  {Yuan} F.,   {Ostriker} J.~P.,  2021, \mn@doi [\mnras]
  {10.1093/mnras/staa3755}, \href
  {https://ui.adsabs.harvard.edu/abs/2021MNRAS.501..398Y} {501, 398}

\bibitem[\protect\citeauthoryear{{Yoon}, {Yuan}, {Ostriker}, {Ciotti}  \&
  {Zhu}}{{Yoon} et~al.}{2019}]{Yoon+19}
{Yoon} D.,  {Yuan} F.,  {Ostriker} J.~P.,  {Ciotti} L.,   {Zhu} B.,  2019,
  \mn@doi [\apj] {10.3847/1538-4357/ab45e8}, \href
  {https://ui.adsabs.harvard.edu/abs/2019ApJ...885...16Y} {885, 16}

\bibitem[\protect\citeauthoryear{{Yuan} \& {Narayan}}{{Yuan} \&
  {Narayan}}{2014}]{Yuan2014}
{Yuan} F.,  {Narayan} R.,  2014, \mn@doi [\araa]
  {10.1146/annurev-astro-082812-141003}, \href
  {https://ui.adsabs.harvard.edu/abs/2014ARA&A..52..529Y} {52, 529}

\bibitem[\protect\citeauthoryear{{Yuan}, {Bu}  \& {Wu}}{{Yuan}
  et~al.}{2012}]{Yuan2012}
{Yuan} F.,  {Bu} D.,   {Wu} M.,  2012, \mn@doi [\apj]
  {10.1088/0004-637X/761/2/130}, \href
  {https://ui.adsabs.harvard.edu/abs/2012ApJ...761..130Y} {761, 130}

\bibitem[\protect\citeauthoryear{{Yuan}, {Gan}, {Narayan}, {Sadowski}, {Bu}  \&
  {Bai}}{{Yuan} et~al.}{2015}]{Yuan2015}
{Yuan} F.,  {Gan} Z.,  {Narayan} R.,  {Sadowski} A.,  {Bu} D.,   {Bai} X.-N.,
  2015, \mn@doi [\apj] {10.1088/0004-637X/804/2/101}, \href
  {https://ui.adsabs.harvard.edu/abs/2015ApJ...804..101Y} {804, 101}

\bibitem[\protect\citeauthoryear{{Yuan}, {Ostriker}, {Yoon}, {Li}, {Ciotti},
  {Gan}, {Ho}  \& {Guo}}{{Yuan} et~al.}{2018a}]{Yuan+20}
{Yuan} F.,  {Ostriker} J.~P.,  {Yoon} D.,  {Li} Y.-P.,  {Ciotti} L.,  {Gan}
  Z.-M.,  {Ho} L.~C.,   {Guo} F.,  2018a, arXiv e-prints, \href
  {https://ui.adsabs.harvard.edu/abs/2018arXiv180705488Y} {p. arXiv:1807.05488}

\bibitem[\protect\citeauthoryear{{Yuan}, {Yoon}, {Li}, {Gan}, {Ho}  \&
  {Guo}}{{Yuan} et~al.}{2018b}]{Yuan+18}
{Yuan} F.,  {Yoon} D.,  {Li} Y.-P.,  {Gan} Z.-M.,  {Ho} L.~C.,   {Guo} F.,
  2018b, \mn@doi [\apj] {10.3847/1538-4357/aab8f8}, \href
  {https://ui.adsabs.harvard.edu/abs/2018ApJ...857..121Y} {857, 121}

\makeatother
\end{thebibliography}



\appendix

\section{Metallicity Distribution}
\begin{figure}
	\includegraphics[width=\columnwidth]{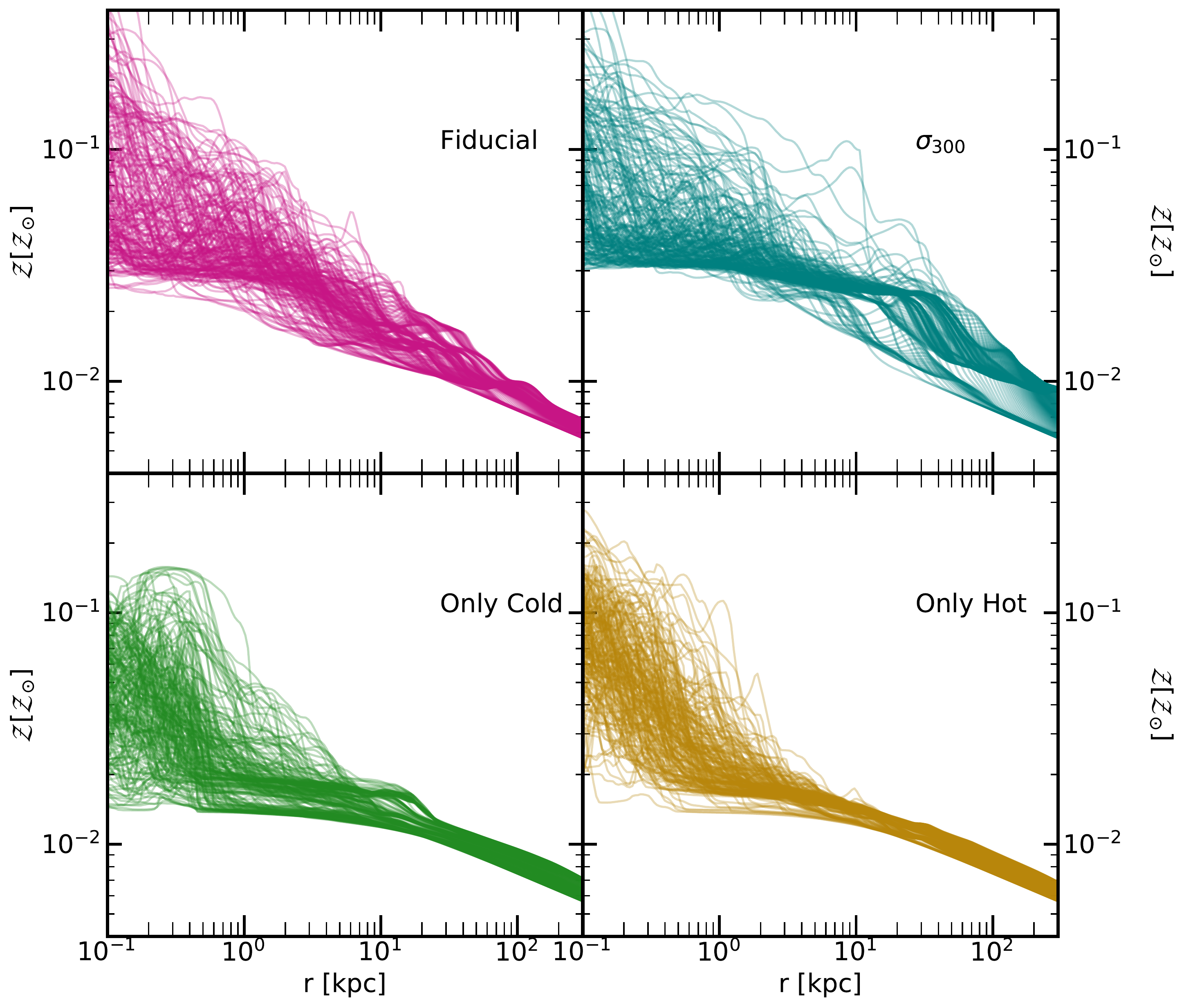}
    \caption{Radial dependence of the metallicity.}
    \label{fig:Zmet_rad}
\end{figure}

\begin{figure}
	\includegraphics[width=\columnwidth]{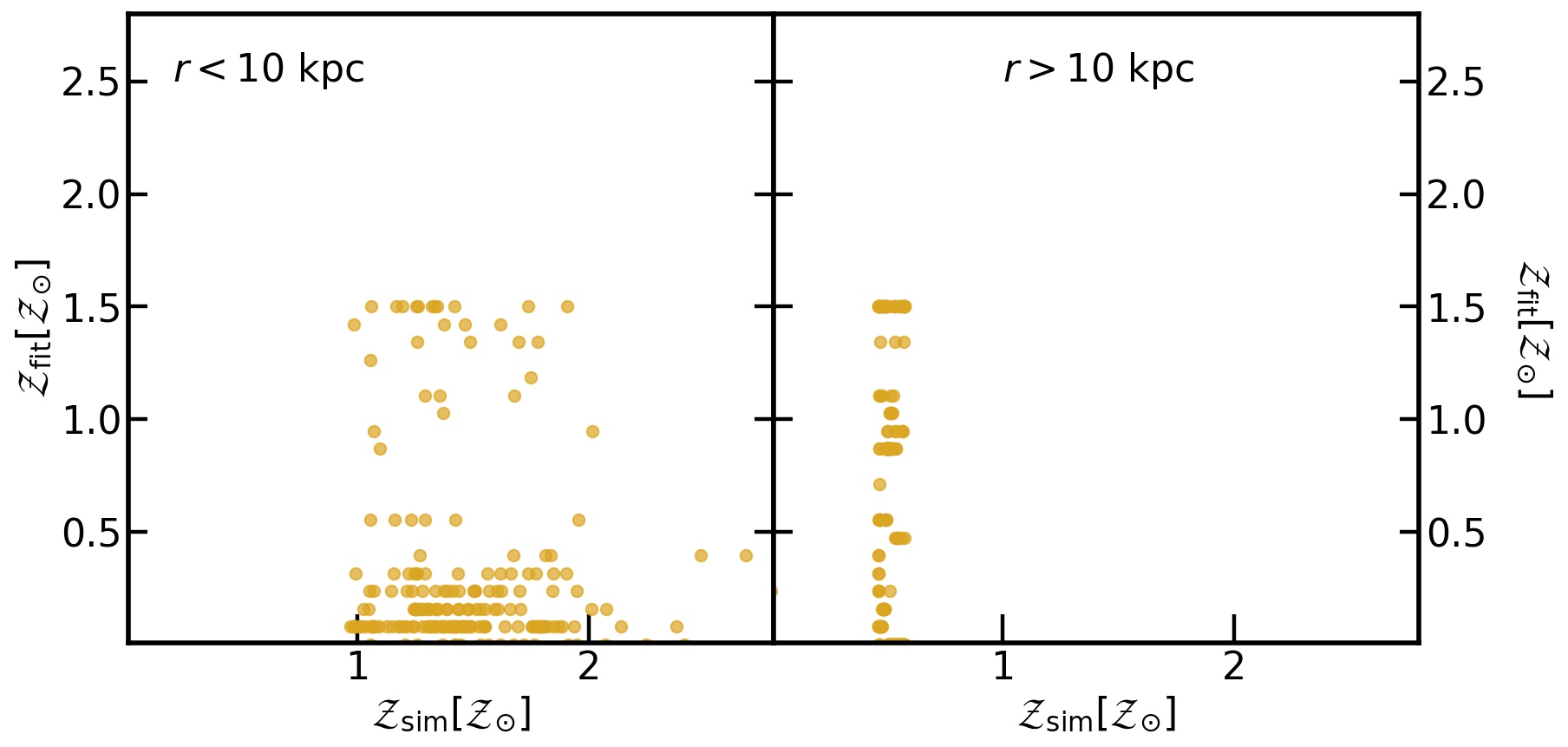}
    \caption{The distribution of the metallicities obtained by fitting the spectra with the volume-weighted metallicities from the simulation data.}
    \label{fig:Zsim_zfit}
\end{figure}

Apart from the density and temperature of the gas, its metallicity also affects the emission spectrum. In Figure \ref{fig:Zmet_rad}, we show the radially averaged metallicity profiles for the four models. As in Figures \ref{fig:tew_vs_rad} and \ref{fig:nH_vs_rad}, each curve represents a different time step and the black dashed curve is the temporal average. Similar to the density and temperature profiles, there is significant variation in the region close to the centre, while at a larger radius the metallicity drops by nearly two orders and does not vary much with time.  

We can extend the analysis discussed in Section \ref{sec:extract_pure_spectra} to test the reliability of the 1-T spectrum fitting paradigm in retrieving abundance information from the spectra. We use the metallicity information provided in pyatomDB tables to extract the gas abundances by following the spectrum fitting procedure outlined for obtaining \tspec\ from the spectra. We follow this exercise for the ISM and the CGM regions of the Fiducial run for every time step of the simulation. We show the variation of the fitted gas metallicities from the spectra ($\mathcal{Z}_{\rm fit}$) with the volume-weighted metallicities ($\mathcal{Z}_{\rm sim}$) obtained from ISM (left) and the CGM (right) regions of the simulation data. The discreteness in $\mathcal{Z}_{\rm fit}$ is representative of the underlying discreteness of the pyatomDB database. For both the regions $\mathcal{Z}_{\rm fit}$ differs considerably from the actual value derived from the simulation. The difference between $\mathcal{Z}_{\rm fit}$ and $\mathcal{Z}_{\rm sim}$ is especially stark in the CGM region where $\mathcal{Z}_{\rm sim}$ does not change significantly from $\sim 0.5 \mathcal{Z}_{\odot}$. We conclude from this Figure that the 1-T model, as described in this paper, cannot reliably retrieve abundance information. 

\section{Comparing 2D and 3D Data}

\begin{figure}
	\includegraphics[width=\columnwidth]{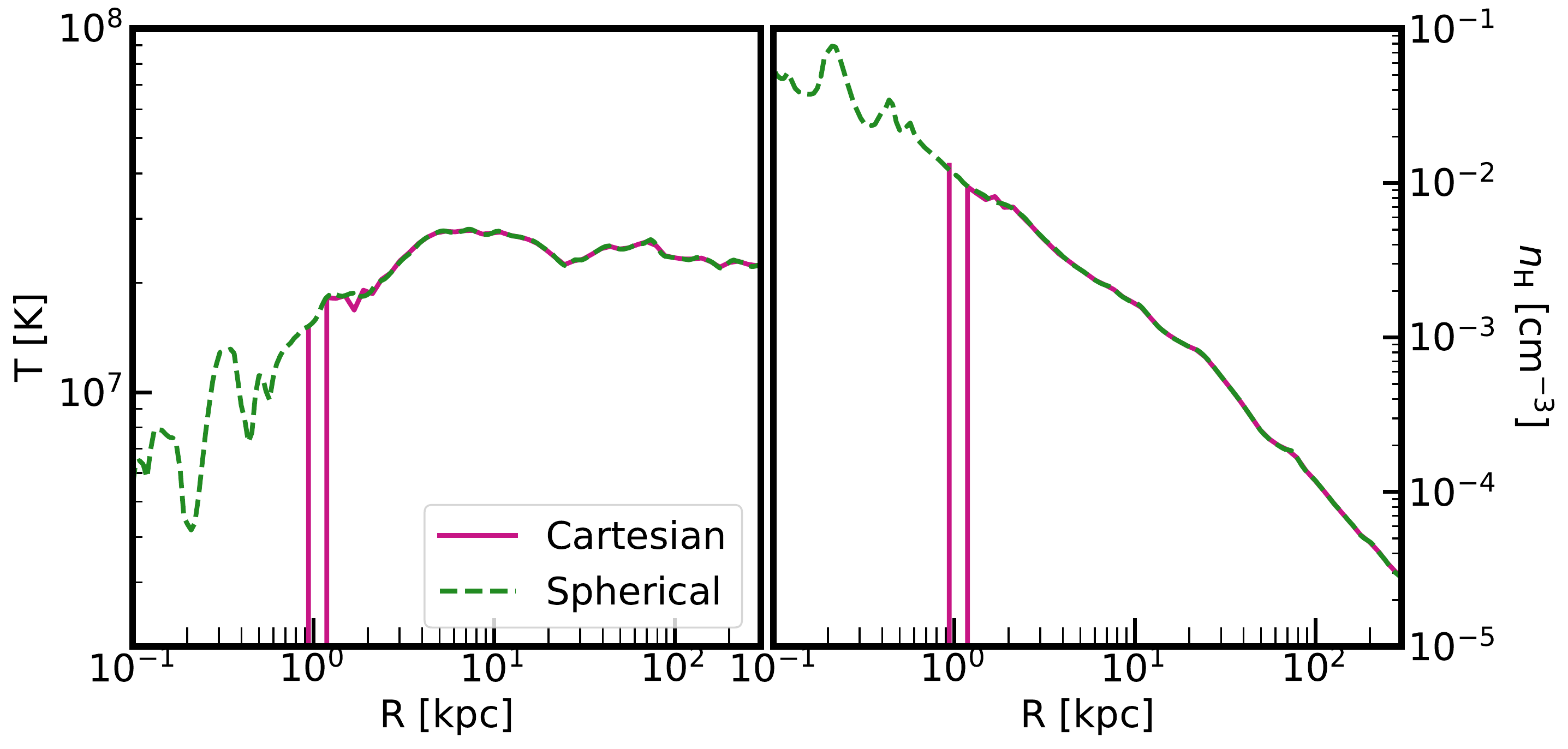}
    \caption{Comparison of the density and temperature profiles from the cartesian and spherical data.}
    \label{fig:dens-temp-comparison}
\end{figure}

We have relied on generating 3D data from the axi-symmetric 2D simulations for the purpose of producing mock images and the radial surface brightness profiles in Section \ref{sec:Mock_Image}. In this Section, we compare the 3D and the original 2D data sets for the Fiducial run. 

In the 2D simulations, the resolution is very high close to the centre ($\sim 0.5\ \rm{pc}$) and decreases to ($\sim 10\ \rm{kpc}$) at larger radius. As we are interested in the diffuse emission from CGM, for the 3D data conversion, we use a uniform grid size of $750$ pc across the domain. In the inner regions of the 2D data, where the resolution is lesser than $750$ pc, we average the quantities over multiple cells in the 2D data set for the corresponding spatial coordinates in the 3D data set. For the rest of cells in the 2D data set, we choose the value of the quantity in the nearest neighbour to the corresponding spatial location in the 3D.

To understand if the conversion between 2D and 3D data was faithful, we compare radially averaged temperature (left) and density (right) profiles for a particular timestep for the Fiducial run. The dip in the profiles, at around $\sim 1$ kpc, indicate the location of the first cell from the centre for the 3D data set. Beyond this radius, the profiles from 2D spherical data and 3D cartesian data are identical.

\section{The Effect of Foreground \& Background on Galaxy Spectrum}\label{sec:fgbk_appendix}

We discuss in brief the foreground and background models added to generate the mock image in Figure \ref{fig:mock_image}. SOXS provides two foreground models, a point source background model and an instrumental background model. For Figure \ref{fig:mock_image}, we rely on the foreground model discussed in \cite{McCammon+2002} which assumes that the foreground is a result of emission from two components- the hot halo at $2.611 \times 10^6$ K and the local bubble at about a $\sim 1 \times 10^6$ K. The point source background model is taken from \cite{Lehmer+2012} and it ranges between $7.63\times 10^{-22}- 1.0\times10^{-12}$ erg s$^{-1}$ cm$^{-2}$.

To understand the effect of foreground and background emission on the spectra, switch on the contribution from foreground and background in SOXS while computing the source spectrum from the source galaxy. We follow the fitting procedure described in Section \ref{sec:extract_pure_spectra} and fit this spectrum with a 1-T model. The fitting temperature is $2.615 \times 10^6$ K, which is very close to the temperature of the hot halo in the foreground. This temperature is much lower than the fitting temperature obtained from spectrum without foreground/background contribution- $2.2\times 10^7$ K. The foreground emission dominates the emission from the source because HUBS has a large FoV. An observation, having a comparable exposure time of $300$ ks, from a telescope with a smaller FoV will result in a smaller foreground contribution (see Figure $13$ of \cite{Vijayan&Li22}).

To quantify the differences between the signal from the galaxy and the foreground, we estimate the fluxes from the two sources. The X-ray luminosity of the inner $20$ kpc region,the brightest region of Figure \ref{fig:mock_image}, is $2.5 \times 10^{39}$ erg s$^{-1}$. At the distance of $15$ Mpc, this translates into a flux of $\sim 8.5 \times 10^{-13}$ erg s$^{-1}$ cm$^{-2}$. We can use SOXS to generate the spectrum from foreground and background. By assuming that the HUBS detector has an effective area of $500$ cm$^{2}$ \citep{Cui+20}, we estimate the flux from the foreground and background to be $\sim 8.6\times 10^{-12}$, which is nearly an order of magnitude larger than that from the galaxy. Thus, the galaxy spectrum can be studied only after properly subtracting the foreground and background contribution from the observed spectrum.


\bsp	
\label{lastpage}
\end{document}